# SUPERCONDUCTING RF TECHNOLOGY R&D FOR FUTURE ACCELERATOR APPLICATIONS

CHARLES E. REECE and GIANLUIGI CIOVATI

*SRF R&D Department, Thomas Jefferson National Accelerator Facility*
*12000 Jefferson Avenue*
*Newport News, Virginia 23606, USA*
*reece@jlab.org, gciovati@jlab.org*

Superconducting rf technology (SRF) is evolving rapidly as are its applications. While there is active exploitation of what one may term the current state-of-the-practice, there is also rapid progress expanding in several dimensions the accessible and useful parameter space. While state-of-the-art performance sometimes outpaces thorough understanding, the improving scientific understanding from active SRF research is clarifying routes to obtain optimum performance from present materials and opening avenues beyond the standard bulk niobium. The improving technical basis understanding is enabling process engineering to both improve performance confidence and reliability and also unit implementation costs. Increasing confidence in the technology enables the engineering of new creative application designs. We attempt to survey this landscape to highlight the potential for future accelerator applications.

*Keywords*: superconducting rf, niobium, accelerator applications

## 1. Introduction

The press for ever more powerful, economic, efficient, and versatile particle accelerators continues to motivate pursuit of innovative solutions involving superconducting materials, particularly those which can support high electromagnetic fields within rf accelerating structures. The superconducting rf technology (SRF) which fills this need now offers robust solutions within a particular parameter space and offers prospects for yet significantly improved cost per unit energy gain. Such gains make credible a variety of new types of accelerator application solutions.

Other contributions to this volume address current operating systems and rf structure design elements in detail. Here, we provide a high-level characterization of the technological domain that is confidently deployable as demonstrated by operating systems. We follow this with a summary of best demonstrated performance of serviceable SRF devices, although such performance may not yet be attained with desirable confidence or affordable cost. Next we review current research that seeks to answer key questions that stand in the way of maximal exploitation of SRF for particle accelerators, questions regarding optimal niobium material characteristics, performance-limiting mechanisms, chemical and mechanical forming and treatment processes, and other material systems that may someday surpass standard bulk niobium with either performance and/or cost effectiveness. Finally, we briefly review candidate future applications of SRF technology that are at various stages of conception.

## 2. Summary of the "State-of-the-Practice" for SRF-based Accelerator Technology

The current "state-of-the-practice" of SRF technology applied to accelerators is represented by those performance levels established as specifications for funded and launched projects and this on the foundation of long-running installations.

The European-XFEL project is fully underway. The performance levels required of this facility have been well demonstrated. The 928 cavities of the full design will operate at 23.6 MV/m accelerating gradient in pulsed mode, 1.4 ms with 10 Hz repetition rate, with anticipated $Q_0$ of $10^{10}$ at 2.0 K [1, 2].

The largest current implementation of SRF technology is the CEBAF recirculating electron linac at Jefferson Lab, with over 350 operational cavities. This facility is presently installing a major upgrade to 12 GeV, based on the addition of ten new high-performance cryomodules [3, 4]. During the final week of physics run in the 6 GeV era, one of the new upgrade cryomodules demonstrated full design requirements of 460 μA CW, providing 108 MeV gain per pass [5]. The eight 7-cell cavities provided the project-specified average 19.2 MV/m with a $Q_0$ of $>8 \times 10^9$ at 2.07 K.

The ISAC-II accelerator at TRIUMF is operational with 40 quarter-wave resonators (QWR) of three types. The ISAC-II performance goal is to operate CW at a gradient of 6 MV/m, corresponding to a peak surface field of 30 MV/m and peak magnetic field of 60 mT with a cavity power ≤7 W [6]. While the heavy ion accelerator ALPI at INFN-Legnaro sustains operation of 64 QWRs running at 3-6 MV/m using a mix of SRF material technologies [7].

The SPIRAL 2 accelerator at GANIL has twelve $\beta = 0.07$ and fourteen $\beta = 0.12$ 88 MHz QWRs operating at 4.2 K. The cavities operate in CW at a gradient of 6.5 MV/m with less than 10 W dissipated power per cavity. During vertical tests, the cavities achieved peak surface electric fields up to ~56 MV/m and peak surface magnetic fields up to ~100 mT with Q-values of ~$1 \times 10^9$ at 4.2 K.

The sixteen 400 MHz LHC SRF cavities are running now, each capable of providing up to 2 MV for beam operations with 5.5 MV/m, having been tested to 8 MV/m [8, 9].

SRF cavities are serving dutifully in electron storage ring applications around the world. In such high beam loading applications, the chief challenges tend to be input power couplers and convenient damping of higher-order-mode (HOM) power in contrast to SRF cavity gradient and $Q$ [10].

The ATLAS Energy Upgrade cryomodule containing seven $\beta = 0.15$ QWRs cavities operating at 109 MHz is operating with beam, providing 14.5 MV of accelerating voltage in 4.5 meters. This is a record for this beta range and represents a factor of three performance gain over prior ATLAS technology [11].

The seven cryomodules of FLASH at DESY are operating reliably at 20-25 MV/m with 1.4 ms pulses, 10 Hz rep rate [12].

A third-harmonic cryomodule for phase space linearization was contributed by FNAL to FLASH. It contains four 3.9 GHz SRF cavities produced by US members of the Tesla Technology Collaboration (TTC). Performance of these cavities at DESY comfortably exceeds the design gradient of 14 MV/m, with each being clean of field emission-induced x-rays to 20 MV/m and quench fields greater than 24 MV/m [13].

The Spallation Neutron Source (SNS) has seven years of commissioning and operation experience, now delivering 1 MW of beam power on target with 80 SRF 805 MHz cavities. This was the world's first superconducting linac for pulsed proton beams. Although the design gradients were 10.2 MV/m and 15.9 MV/m for the $\beta = 0.61$ and $\beta = 0.81$ cavities, respectively, the operational gradient distributions are essentially the same for both sets: ~10–15 MV/m, with the dominant performance limitation being field emission-derived heating effects [14].

## 3. SRF "State-of-the-Art" in 2012

### 3.1. *Best β=1 structure performances*

The technologically most demanding SRF challenge recently has been the R&D push to make an International Linear Collider (ILC) technically viable and credibly affordable. Such a facility would dwarf all other SRF applications in scale. The push to make 35 MV/m confidently attainable has met with significant success. The challenge is principally one of quality control in materials and processes, then secondarily one of cost optimization by process and material optimization.

The international R&D collaboration has yielded several multi-cell cavities with performance exceeding 40 MV/m, and yield of 90% with performance > 35 MV/m has been demonstrated by at least one laboratory [15-17]. The present best 9-cell gradient performance reported has been from two cavities fabricated from ingot Nb by Research Instruments GmbH. These cavities were quench limited in the 25-29 MV/m range after standard acid etching treatment, but quench fields increased to 31-45 MV/m after subsequent electropolishing, which corresponds to peak magnetic surface fields of up to 192 mT [18].

One of the CEBAF upgrade prototype 7-cell cavities, LL002, a fine-grain, high RRR Nb with a history of >250 μm etch removal, reached 43.5 MV/m in a 2010 test, limited by available rf power, after a 34 μm electropolish. For this "low loss" cell shape design, this corresponds to peak magnetic surface fields of 162 mT. One of the few 12 GeV Upgrade production cavities actually tested to its limits, C100-RI-006 quenched at 41.6 MV/m (Figure 1).

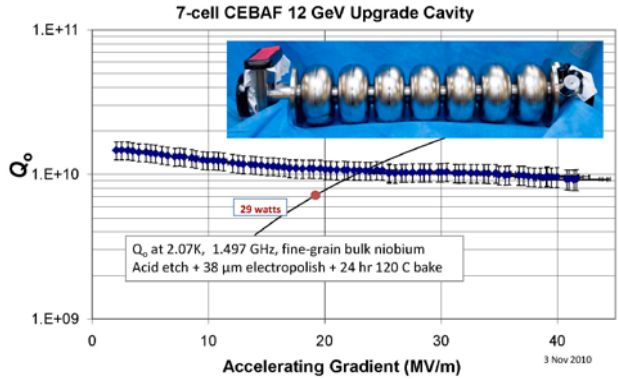

Figure 1: Best CEBAF Upgrade cavity performance test.

### 3.2. *Best performance from β<1 structures*

Investments in careful process analysis and engineering are also yielding performance progress in Nb cavities with the more complicated geometries required for β<1 applications. ANL recently set a new record for accelerating gradient with a 72 MHz β=0.077: $E_{acc}$ =12 MV/m and $Q_0$ of $6×10^9$ [19].

Similarly, the revised designs of two 80.5 MHz QWRs and two 322 MHz half-wave resonators (HWRs) required for FRIB have recently been tested and exceeded all requirements. The residual resistance measured in the three prototype families was below 5 nΩ up to about 100 mT in QWRs, and about 80 mT in HWRs [20].

FNAL is developing a β=0.22 325 MHz single spoke resonator (SSR1) for a possible Project X front end. A recent test demonstrated the project requirement of $E_{acc}$ =12 MV/m and $Q_0$ of $5×10^9$ [21].

## 4. Current SRF Research

### 4.1. *Bulk Nb material R&D*

The technical specifications for bulk Nb material commonly used for the fabrication of SRF cavities are listed in Table 1 [22]. Parameters other than RRR and impurity content are important for good formability of the material. The RRR value has been specified in order to assure good thermal conductivity, κ, of the material, as κ at 4.2 K is directly proportional to RRR. High thermal conductivity allows for thermal stabilization of defects. A simplified model shows that the quench field, $H_{quench}$, due to a normal-conducting defect of surface resistance $R_n$ and radius $a$ is proportional to the square root of κ [23]:

$$H_{quench} \propto \sqrt{\frac{\kappa}{a R_n}} . \qquad (1)$$

Specifications for the content of main impurities in Nb were dictated by the need to assure a high RRR value and to limit the concentration of impurities such as hydrogen and oxygen which have well known negative impact on the superconducting properties of Nb [24, 25]. The understanding of various factors which contribute to Nb RRR continues to mature [26].

Table 1. Specifications for Nb bulk material used for the fabrication of SRF cavities.

| | |
|---|---|
| RRR | > 300 |
| Grain size ($\mu$m) | $\approx$ 50 |
| Yield strength, $\sigma_{0.2}$ (MPa) | $50 < \sigma_{0.2} < 100$ |
| Tensile strength | > 100 |
| Elongation at fracture | 30% |
| Vickers hardness | $\leq$ 50 |
| Content of main impurities (wt.ppm) | Ta $\leq$ 500; O $\leq$ 10; N $\leq$ 10; C $\leq$ 10; H $\leq$ 2 |

Few companies throughout the world can provide the Nb material with the specifications listed in Table 1, and that at a cost which is significantly higher than that of lower purity Nb and which has increased rather sharply in the past 5 years. Elaborate manufacturing steps which include forging, grinding, rolling and annealing contribute to the cost of fine-grain Nb sheets for SRF cavity applications. In addition, each of these manufacturing steps can potentially introduce foreign inclusions in the material, which can subsequently result in reduced cavity quench field. Quality control methods, such as the use of eddy current scanning systems to detect inclusions of size of the order of 50-100 $\mu$m have been implemented by some laboratories to inspect Nb sheets received from the manufacturer [22].

In addition to cost, the availability of low-Ta ingot and the sheet niobium production throughput by experienced industrial vendors impose bottlenecks for producing large quantities of niobium sheets envisioned large projects. In the following section we will discuss a new type of material, referred to as "ingot Nb" which has proven to be a viable alternative to the standard "fine-grain" Nb.

### 4.1.1. *Ingot Nb*

In 2005, a collaboration between CBMM, Brazil, and Jefferson Lab [27] demonstrated that the mechanical properties of niobium directly sliced from an ingot were adequate for forming into cavity half-cells and that cavities built from such material achieved performance levels comparable to those of cavities built from standard fine-grain Nb [28, 29].

Nb discs are typically sliced from an ingot to the required thickness by wire electrodischarge machining and have a non-uniform crystal structure, with grains of typical size ranging from few to several centimeters. Studies at DESY proved that the cavity cell shape could be kept within specified tolerances by proper tooling during forming and machining of the half-cells, in spite of the anisotropic mechanical properties of the discs [30].

Sharp steps at grain boundaries, which also result from non-uniform mechanical properties, can be easily eliminated by local grinding or centrifugal barrel polishing. (See section 4.3.1 below.)

The ingot Nb technology for SRF cavity production was fully proven at DESY, where two 9-cell cavities built from this type of material have been installed in a cryomodule in the FLASH accelerator [30]. In addition, DESY qualified eight additional nine-cell cavities, which will be installed in the first cryomodule for the XFEL project. Several of these cavities reached accelerating gradients greater than 40 MV/m, with one of them establishing a record-setting gradient for these 9-cell TESLA-style cavities of 45 MV/m [18].

Studies at Jefferson Lab and DESY also indicated that 20-30% lower surface resistance was obtained in multi-cell cavities built from ingot Nb, compared to cavities built from fine-grain Nb, for the same treatment procedure, cavity frequency and operating temperature [29, 31, 32]. A reduction of the surface resistance by a factor of ~4 was recently obtained at Jefferson Lab with an ingot Nb single-cell cavity heat treated at high-temperature in a vacuum furnace [33]. Investigation into the specific material characteristics which produce this nice result are underway.

The obvious reduction of the cost of fabrication of ingot Nb discs, along with the reduced need for their quality control compared to fine-grain Nb sheets, makes them a cost-effective solution for the fabrication of SRF cavities. Many companies worldwide are able to offer ingot Nb material and the introduction of multi-wire slicing techniques makes the mass-production of ingot Nb discs possible. The above mentioned evidence for better performance of ingot Nb cavities, compared to fine-grain cavities, which has emerged in the last few years adds to the attractiveness of ingot Nb material for SRF cavity fabrication.

Current R&D topics related to ingot Nb material include measurements of superconducting, thermal and mechanical properties on samples. Measurements of DC critical fields and magnetization done at Jefferson Lab indicated a lower pinning current density for ingot Nb samples than fine-grain ones [34]. This result, together with the reduced flux trapping efficiency of ingot Nb compared to fine-grain Nb, as it was measured on samples at BESSY [35], might explain the lower RF residual resistance of ingot Nb cavities.

Studies on the mechanical properties of ingot Nb are discussed in Sec. 4.1.4. Low-temperature thermal conductivity measurements done on ingot Nb samples at several laboratories showed that, unlike for fine-grain Nb where $\kappa$ is limited at $\sim 2$ K by phonon scattering at grain boundaries, $\kappa$ has a pronounced "phonon peak" at $\sim 2$ K [36-38]. Although the phonon peak is reduced by plastic deformation of the samples, it can be restored by a subsequent high temperature heat treatment [36]. This result would imply better thermal stabilization of the cavity inner surface which results in a reduced dependence of the surface resistance with increasing RF field, which is often observed for such heat treated ingot Nb cavities at 2 K.

### 4.1.2. *Single-crystal Nb*

Ingot Nb discs with a large ($\sim 20$ cm) diameter central single-crystal allow the fabrication of single-crystal cavities. Two $\sim 2.3$ GHz single-cell prototypes were built and tested at Jefferson Lab and both reached a peak surface magnetic quench field of $\sim 160$ mT at 2.0 K [39].

A method based on rolling and annealing steps, was developed at DESY to enlarge a single-crystal disc to a diameter of $\sim 23$ cm, suitable for 1.3 GHz cavity fabrication, while maintaining a single-crystal structure [40]. It was also shown that, by properly matching the orientation of two half-cells, the single-crystal structure was preserved after electron-beam welding the half-cells. RF tests of 1.3 GHz single-cell prototypes at 2.0 K reached quench field values of $\sim 160$ mT [41].

The main attractiveness of single-crystal cavities over typical ingot Nb is the uniformity of mechanical properties and therefore better forming. So far, there seems to be no significant cavity performance improvement compared to cavities made of "standard" ingot Nb, with few cm-sized grains.

Efforts to produce large-diameter single-crystal Nb ingots have been pursued by KEK in collaboration with Tokyo Denkai in Japan [42].

### 4.1.3. *Niobium purity and tantalum content*

As mentioned in Sec. 4.1 above, the requirement of high purity (RRR > 300) Nb was dictated by the need for high thermal conductivity to avoid premature quenches. A correlation between higher $\kappa$ and increased quench field was indeed noticeable when looking at a data set of nine-cell cavities measured at DESY and made of fine-grain Nb, treated by BCP. Nevertheless, when including data from nine-cell cavities made of fine-grain Nb and treated by electropolishing, the correlation between $\kappa$ and $H_{quench}$ is lost, as shown in Fig. 5.61 in Ref. [43], for RRR values ranging from $\sim 200$ to $\sim 750$. The same conclusion was found at KEK, for RRR values ranging between 130 and 800 [44].

Additional evidence for the reduced impact of high RRR on quench field was provided by ingot Nb cavities: as shown in Fig. 2, there is no dependence of the quench field (corresponding to $E_{acc}$-values > 25 MV/m) on RRR, ranging between 150 and 500, as measured on nine-cell ingot Nb cavities treated by BCP [30].

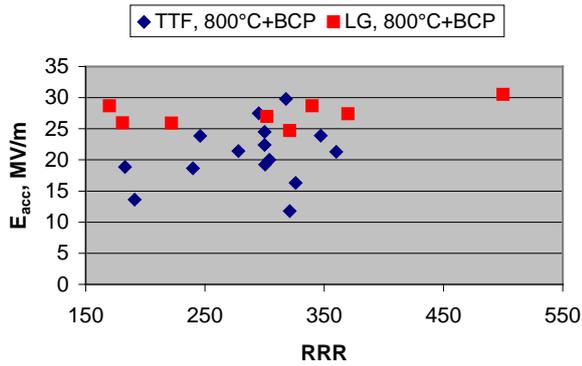

Fig. 2. Maximum accelerating gradient as a function of RRR for 1.3 GHz nine-cell cavities made of fine-grain (blue diamond) and ingot Nb (red squares), heat treated at 800°C and etched by BCP. The figure was taken from [30].

These findings suggest that normal conducting inclusions of size of the order of tens of micrometers are not anymore the dominant cause of quench in SRF cavities.

In the case of ingot Nb material, the possible presence of a phonon peak at ~ 2 K breaks the connection between thermal conductivity and RRR. As it is shown in Fig. 3, for example, single-crystal samples of lower RRR-values have higher κ(2K) than fine-grain samples of higher RRR. As mentioned before, high temperature heat treatments provide a method to enhance κ(2K) after forming.

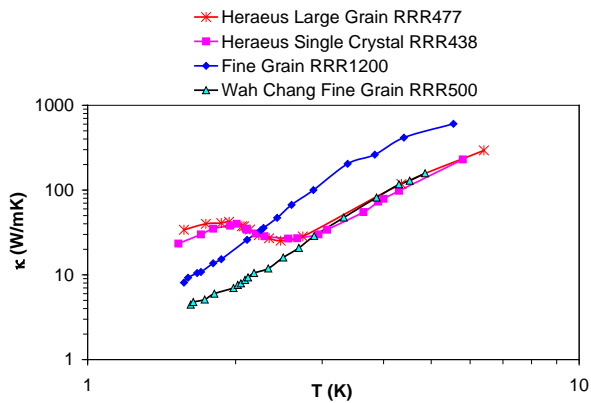

Fig. 3. Thermal conductivity as a function of temperature of fine-grain and ingot Nb (large-grain or single-crystal) samples. The figure was adapted from [36].

These recent findings suggest that, for cavities with $E_{acc}$-specification below ~30 MV/m, Nb material with RRR greater than ~150 is acceptable. Additional accumulated experience with cavities built from this "medium-purity" material could further strengthen this statement.

Tantalum is a "substitutional" impurity in Nb and is therefore difficult to separate from the Nb itself. In 1996 it was found that a premature quench in a cavity was due to a specific Ta cluster [45]. In order to reduce the possibility of such "clusters" being present in the Nb sheets, the concentration of Ta in Nb was specified to be ≤ 500 wt. ppm. This specification imposes a significant increase in the cost of the Nb sheets and precludes the use of Nb extracted from pyrochlore ore, which is a significant fraction of the sources of Nb production. The mechanism for creating such clusters and their frequency of occurrence is unclear.

Studies on single-cell cavities and samples have been done in recent years to re-evaluate the role of Ta. Magnetization measurements show no significant difference among ingot Nb samples with Ta content between 600-1300 wt. ppm [31, 33, 36, 46]. Tests of single-cell cavities made of fine-grain Nb showed no correlation between Ta content, between 600-1300 wt. ppm, and maximum $E_{acc}$ to ~30 MV/m [47]. Peak surface magnetic field quench values of ~100 mT have been obtained in multi-cell cavities made of ingot Nb with Ta content up to 1500 wt. ppm [30].

These results suggest that the Ta content specification can also be relaxed without significant loss in cavity performance. If the presence of Ta clusters is feared, eddy current scanning can be used to exclude Nb discs with such defects from the cavity fabrication process [24].

The potential reduction in the cost of the material which can be obtained by relaxing the RRR and Ta content specifications may be very attractive, particularly for small accelerators with "moderate" gradient specifications for research at universities or for industrial applications. These results suggest that the Ta content specification can also be relaxed for cavities requiring only medium gradients of less than 30 MV/m.

### 4.1.4. *Mechanical properties*

Renewed efforts to better understand the mechanical properties of both fine-grain and ingot Nb and their implications on cavity performance and formability have been undertaken by several laboratories and universities.

One of the final steps for the production of fine-grain Nb sheets is a 2% thickness reduction by rolling. This process causes higher strain to be localized near the surface, smaller grain size and different texture at the surface, compared to deeper in the bulk. Orientation image maps (OIM) obtained by electron backscattered diffraction (EBSD) showed cross-sectional microstructure and texture gradients to vary greatly among samples cut from fine-grain Nb sheets [48]. This results in variability in forming half-cells and in the metallurgical state after high temperature heat treatments or electron beam welding (EBW).

The yield strength of fine-grain niobium decreases up to ~40% after heat treatment up to ~1200°C. The yield strength also depends on purity (lower purity Nb has higher yield strength) and grain size [49]. For example, as-received ingot Nb samples showed yield strength values ranging between 30-50 MPa [45, 50], compared to ~55–80 MPa for as-received high-purity fine-grain Nb [51]. Issues with batches of fine-grain Nb sheets not fully re-crystallized (non-uniform grain size throughout the thickness) have occurred in the past and resulted in unacceptably low yield strength of the material after heat treatment at 800°C [52]. As a result, the heat-treatment temperature for hydrogen degassing for SNS and CEBAF upgrade cavities at Jefferson Lab was 600°C. The 600°C recipe was also used for ILC R&D cavities initially at JLab; due to excessive stiffness issues, a 800°C recipe replaced the 600°C recipe [53]. A 800°C bake recipe has been used for a long time at DESY.

In the past years, there has been increasing evidence for the metallurgical state of the material influencing not only cavity formability and robustness but also the cavity performance:

- A correlation between high dislocation content and enhanced losses at high (>90 mT) RF field has been reported [54].
- A correlation between etch pits, crystal defects and enhanced RF losses at medium (20 mT $<B_p<$ 90 mT) field has been reported [55].
- EBW of cold-worked fine-grain Nb can induce pitting in the zone where thermal stress accumulates (so called "heat affected zone") [56]. It was shown that pits in this zone caused premature quenches in some 1.3 GHz nine-cell cavities [57, 58].

All the above considerations suggest the need for optimization studies to identify the proper sequence of processing steps and heat treatment parameters, for each type of material, which gives the best compromise between mechanical properties and cavity performance. For ingot Nb material, it would be important to identify crystal orientations which don't behave well during forming and should therefore be avoided during the ingot production by the manufacturer. Such study is ongoing at MSU [59, 60].

## 4.2. **Bulk Nb Cavity Fabrication**

The techniques most commonly used for the fabrication of bulk Nb cavities are deep-drawing of Nb sheets and electron-beam welding of the formed parts.

Electron-beam welding of Nb is a critical process which requires machine-specific parameters be mastered in either industry or laboratory setting. Occasionally, issues with defects in the welds or in the heat affected zone limit the cavity performance to $E_{acc}$~15-20 MV/m [30, 54, 55, 61].

Forming processes which could avoid most of the welds in a multi-cell cavity are hydroforming, which has been pursued mostly at DESY, and spinning, which has been used at INFN-Legnaro. Initial studies on single-cell cavities made by either hydroforming [62] or spinning showed that $E_{acc}$~40 MV/m was achievable [63], comparable to the best results obtained from cavities made by standard deep-drawing and welding.

The hydroforming machine at DESY allows the fabrication of up to three-cell 1.3 GHz cavities and three full nine-cell prototypes were built by welding three-cell units. The nine-cell cavities were treated by EP and achieved $E_{acc}$~30-35 MV/m.

A 1.3 GHz nine-cell cavity was also built by spinning [63], but a hole was made in one of the irises after centrifugal barrel polishing (CBP) because of non-uniform thickness of the material.

Both the hydroforming and spinning of cavities require a seamless Nb tube of the appropriate diameter and uniform fine-grain structure, usually made "in-house" by deep-drawing, flow forming, extrusion or a combination of those techniques. The use of "single-crystal" seamless Nb tubes has also been recently proposed [64].

The hydroforming technique was also applied to fabricate cavities of Nb/Cu clad material. They could have the advantages of reduced cost because of less Nb and fewer welds and improved thermal conductivity and stiffness from the backing Cu. Single-cell cavities have been built both at DESY and KEK reaching $E_{acc}$-values of up to 40 MV/m [58]. Temperature gradients across the cavity have to be minimized near $T_c$ in order to avoid trapping magnetic field attributed to thermoelectric currents. Reduction of $Q_0$-values after quenches have been measured and attributed to this same effect [58].

The starting Nb/Cu clad tube had been made by different techniques such as explosion bonding, back extrusion and hot rolling. There is yet no clear indication of a preferred tube forming method, and there hasn't been significant progress towards fabrication of cavities longer than two-cells in the past few years.

## 4.3.  *Surface finishing – for quality control and economy*

A very thin layer of material in the cavity surface determines the response to rf fields. The superconducting penetration depth for Nb is only ~40 nm. Efforts to obtain optimum performance for accelerator applications, then, focus very strongly on controlling the morphology and structure of that thin layer of material.

### 4.3.1.  *Centrifugal barrel polishing*

Initially started as an effort to obtain a standard surface finish and reduce the required quantities of hazardous processing chemicals, the use of centrifugal barrel polishing (CBP) on Nb SRF cavities was reported by Higuchi, Saito et al. at KEK in 1995 [65]. The first batch of single-cell cavities to attain consistent ≥40 MV/m performance were prepared by CBP, followed by light EP [66, 67], in 2006.

Recently the CBP treatment process has been picked up and extended at FNAL and JLab in an effort to produce consistently smooth surfaces, somewhat forgiving of the details of the prior fabrication processes [68, 69]. The goal of this work is to find a reliable "low-tech" and inexpensive process that transforms the high-field surfaces into a smooth, uniform surface that subsequently requires only a minimum of chemical processing to leave a "crystallographically clean" surface exposed. While further optimization is underway, examples have demonstrated that otherwise flawed and otherwise unacceptable cavities can be brought to peak performers via inclusion of CBP in their treatment [70].

### 4.3.2.  *Acid etching of Nb – BCP*

Long the standard way of exposing fresh material in bulk Nb cavities, etching with $HF:HNO_3:H_3PO_4$ reagent acids mixed in ratios of either 1:1:1 or 1:1:2 is commonly referred to as buffered chemical polish, or BCP. The exothermic chemical reaction requires thermal management to obtain uniform results. Nb cavities are typically etched with 15ºC circulating acid. The details of technique for realizing predictable etching removal rates over increasingly complicated geometrical shapes continue to be explored [71].

As the target operational surface fields of SRF structures have been pushed up in recent years, the natural residual surface roughness of fine grain Nb after BCP treatment is found to induce limiting effects, apparently due to local field enhancements that subsequently lead to non-linear losses. Such roughness is attributed to variation in local etch rate of different exposed crystal planes and to residual stresses from dislocations and other defects [55]. The roughness and the associated non-linear losses are greatly reduced in heat-treated ingot Nb surfaces.

### 4.3.3. *Standard electropolishing of Nb*

The highest field performance of Nb SRF cavities so far are those with surface smoothened by electropolish in contrast to etching. While this has now been unambiguously demonstrated for fine-grain Nb, it also appears to be the case for ingot Nb, although the incremental benefit is less and conditions of improvement are less well resolved.

The standard Nb electropolish (EP) that has evolved from the Siemens recipe involves $HF:H_2SO_4$ reagent acids in ration of ~1:9 as the electrolyte, with Nb anode and high-purity Al cathode. The typical physical arrangement mounted first by KEK for the TRISTAN cavities has the cavity and collinear cathode horizontal with the electrolyte filling approximately 60% of the cavity [72]. Recent involvement of electrochemical insights and methods in Nb EP research has yielded a clarified understanding of the polishing mechanism and in turn provided pointers toward process optimization [73-75]. Identification of the importance of managing process temperature control for both uniform removal and optimum surface leveling, together with recognition of the source and methods for reducing and removing precipitating sulfur have borne fruit in increased cavity performance for the ILC R&D effort [17], the JLab 12 GeV Upgrade cavities [76, 77], the ANL QWR cavities [19], and the high-grad cavity set for the XFEL [18], among others.

The 80 CEBAF upgrade cavities, for example, were given a consistent 30 µm temperature-controlled EP following a 160 µm BCP etch provided by the cavity manufacturer. The resulting performance was sufficiently reliable that most of the cavities were given their cryogenic acceptance test only after their helium vessel had been welded on, and testing was administratively limited to 27 MV/m [78]. (No more than 23 MV/m will be useful in CEBAF due to other limitations in the accelerator.)

### 4.3.4. *Vertical electropolish*

While Nb EP yields smoother and in many cases surfaces that can support higher fields than the relatively easy BCP etch process, the commonly implemented methods are rather complex and labor intensive compared with what one would like. Because of this, efforts are underway at several labs to develop cavity EP with the cavity filled with electrolyte and oriented with central axis vertical (VEP). Such an arrangement would have several advantages, not least of which would be its greater prospect for process automation.

Cornell has recently reported successful VEP of a 9-cell Tesla-style cavity with field performance exceeding 35 MV/m [79]. CERN and Saclay are investing in new VEP systems, and JLab has repurposed a closed chemistry cabinet for VEP R&D [80, 81]. In addition to temperature regulation, the vertical orientation presents additional challenges for managing the hydrogen bubbles evolved from the cathode and discerning and providing the appropriate level of electrolyte circulation for predictable, uniform removal.

## 4.4. *Surface cleaning for field emission control*

Surface cleanliness is a persistent requirement of SRF accelerating structures. The best prepared low-loss surface can be made irrelevant by a small amount of contaminating particulates. Such particulates easily become sources of field emission when they find their way to high surface electric field regions of a cavity. Adequately clean Nb surfaces have been demonstrated to be field emission free to ≥150 MV/m [82], supporting the notion that field emission from typical technical Nb surfaces may be attributed to either mechanical damage or extrinsic contamination.

Residue from chemical processing steps must be removed else they concentrate on drying. Generous dilution and removal by ultrapure water rinsing in the context of a controlled clean-room environment is considered necessary. Careful contamination control technique is essential to obtaining and retaining adequately clean high-field surfaces. High pressure (>100 bar) rinsing with ultrapure water (HPR) is routinely applied as a final pre-assembly step. The actual protocols for such HPR have little demonstrated basis, however. Techniques at each different laboratory have evolved via loose empiricism rather than quantitative process development. Water pressure, nozzle geometry, flow rate, nozzle-to-surface distance, surface sweep rate all can reasonably be expected to affect particulate removal efficiency, so the domain is yet ripe for careful study and improvement [83, 84].

Ethanol or methanol rinse is employed at some labs to provide assurance of removing precipitated sulfur, others such as Jefferson Lab rely on ultrasonic agitation with a detergent for this purpose.

$CO_2$ snow has been demonstrated to be highly effective at cleaning Nb surfaces [85]. This has not yet, however, been deployed in a way suitable for multi-cell cavities.

## 4.5. *Thermal treatments*

It has been clearly established empirically that low-temperature (120–160°C) baking of etched or electropolished Nb surfaces in ultra-high vacuum or atmospheric noble gas [86] for several hours (LTB) significantly reduces the high-field $Q$-slope, allowing achievement of $B_p$-values of up to ~190 mT at 2.0 K [87]. In addition to that, LTB of such cavities causes a reduction of $R_{BCS}$ and often an increase of the residual resistance, $R_{res}$, so that a moderate 10-20% increase of $Q_0$ at 2.0 K, at low field, is obtained [88].

LTB protocols vary significantly in duration and hold temperature, with 120°C for 24-48 hours just prior to cold rf testing being most commonly used. Discussion of the theoretical understanding of the phenomenon is addressed in section 5.1 below.

While vacuum bake at 600-800°C of near-final Nb cavities is routinely applied for the reduction of interstitial hydrogen to avoid hydride precipitation and its attendant "$Q$-disease," current research is examining more subtle residual surface composition effects that may be influenced by post-chemistry thermal treatments. [89] Early indications from JLab suggest opportunities for reduction of both residual and BCS rf surface resistance with carefully managed surface chemistry during and immediately following ≥1200°C bake of ingot Nb cavities [33, 90].

## 4.6. *Alternate materials to bulk Nb*

### 4.6.1. *Motivations for alternatives to bulk Nb*

With the real prospect of the community being able, in the not so distant future, to specify preparation methods which will reliably yield nearly the theoretical best performance in both supported fields and resistive losses of Nb, one naturally asks the question how might accelerator systems be fielded yet more efficiently than Nb will allow?

Several dimensions merit consideration. If one could realize the same rf characteristics from a thin layer of Nb (less than 1 μm is required) on a less expensive, easily formable and high thermal conductivity substrate, then significant cost savings may be accessible in materials. Also one may have access to more convenient cooling configurations employing, for example, channel cooling rather than bath cooling, which in turn opens new possibilities in the engineering design of accelerator cryomodules.

Alternatively, if materials with higher $T_c$ and adequately high critical fields could be made to realize their apparent theoretical potential, then huge savings in capital and operational cryogenic costs might be realized. $Nb_3Sn$, $NbTiN$, and $MgB_2$ are among the materials receiving such attention.

Further down the road there remains the potential prospect of hybrid multi-layer films which may support dramatically higher fields with very low losses.

### 4.6.2. *Magnetron sputtered Nb films*

The largest fielded application of sputtered Nb film was on the 352 MHz LEP2 cavities. For 272 of these cavities, Nb film on copper was chosen to save cost. These cavities fulfilled their mission for LEP2, but were very hard pressed to perform with adequate $Q$ at 7 MV/m [91].

The best demonstrated performance of Nb-film technology cavities dates from the post-LEP development work for LHC cavities at CERN [92, 93]. 1.5 GHz cavities demonstrated 15 MV/m with $Q_0$ of $10^{10}$ at 1.7 K. The magnetron sputtered cavities show a characteristic non-linear loss. Such limiting $Q$-slope persists in the operating LHC cavities today [8]. The RRR of such sputtered Nb films was found to be ~30.

Nb/Cu medium-$\beta$ QWR cavities have recently been developed for ALPI [94] , and development of sputtered Nb/Cu 101 MHz QWRs for HIE-ISOLDE is underway at CERN [95].

### 4.6.3. *Energetic condensation of Nb*

When considering the gas phase deposition of niobium, it may be helpful to recognize that its very high melting temperature implies very low mobility of atoms near the growth surface at temperatures tolerable to desired application substrates such as copper and aluminum. To realize high crystallinity with acceptably low defect density, additional local energy is required. "Energetic condensation" provides this as kinetic energy of the arriving ions.

A collaboration led by JLab has been characterizing Nb film growth on a variety of model substrates as a function of substrate preparation and energy of arriving Nb ions [96-105]. The methods of creating the energetic ions vary from ECR plasma biased extraction, to cathodic arc discharge plasma, to high power impulse magnetron sputtering (HiPIMS) [106] championed by Andre Anders of LBNL.

Nb films grown by energetic condensation show epitaxial character with low defect density, as measured by RRR quite comparable to that of high-purity bulk Nb now used for cavity fabrication (200–400). While attaining maximum RRR is not itself the goal of deposited Nb films, learning how high-quality Nb crystals develop under controlled conditions on well-known substrates gives insight into tailoring growth conditions for eventual technical applications. Candidate growth conditions have been identified which yield very attractive Nb films grown on fine-grain Cu, such as would make for a fine cavity substrate. Attempts to realize such conditions on test elliptical cavities are planned in the coming year.

### 4.6.4. *Nb₃Sn*

$Nb_3Sn$ continues to receive low-level attention as a material which might exceed the performance of Nb. With a $T_c$ ~18.2 K there is the attractive prospect that high frequency structures made with $Nb_3Sn$ might have negligible losses at the relatively more convenient operating temperature of 4.5 K. Because $Nb_3Sn$ is very brittle and has low thermal conductivity, it must be formed in shape as a film on a suitable substrate. To obtain good stoichiometry from gas phase supplied Sn diffusion into Nb, a reaction temperature above 930ºC is required.

The most thorough work on $Nb_3Sn$ for SRF was carried out at Universität Wuppertal in the 1980's and 90's [107]. Recently work with this material has resumed at Cornell University [108]. Careful material science is yet required to understand and confidently control the $Nb_3Sn$ crystal growth dynamics so as to produce low-loss surfaces to fields corresponding to $E_{acc} > 15$ MV/m.

### 4.6.5. *NbTiN*

Although A15 compounds such as $Nb_3Sn$ have a higher $T_c$, the Nb B1 compounds are less sensitive to radiation damage and crystalline disorder. NbN and NbTiN are the B1-compounds with the highest critical temperature, respectively 17.3 K and 17.8 K. The ternary nitride NbTiN presents all the advantages of NbN and exhibits increased metallic electrical conduction properties with higher titanium percentage. Ti is a good nitrogen getter, so the higher the Ti composition, the lower the number of vacancies. For these reasons, work at Jefferson Lab is targeting NbTiN together with the dielectric AlN for development of SRF multilayer structures [109, 110]. Figure 4 shows a TEM image of a FIB-cut cross-section of a recent multilayer film formed by DC reactive sputtering.

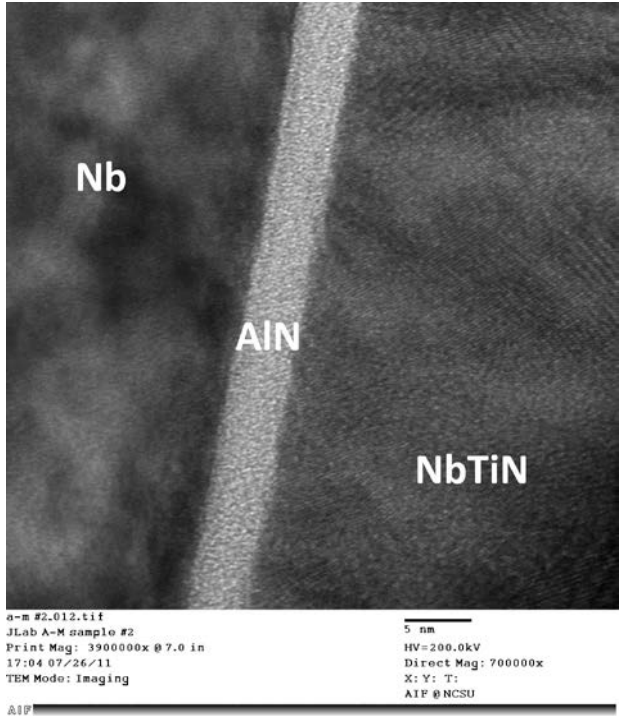

Fig. 4. TEM image of a FIB cut cross-section Cu/Nb/AlN/NbTiN film structure formed by DC reactive sputtering.

### 4.6.6. *MgB₂*

$MgB_2$ is a binary compound that contains hexagonal boron layers separated by close-packed magnesium layers. It has a high critical temperature, $T_c$, around ~39-40 K.

In SRF applications, a $MgB_2$ film with large grain size (>100 μm) is preferred because grain boundaries in the film could cause strong pinning which would contribute to high field Q drop [111]. Rf penetration depth is 100–200 nm at temperatures below 5 K in the GHz range. So a film with several hundred nm thickness is required. Compared to Nb, $MgB_2$ has a smaller lower critical field $H_{c1}$ and a larger upper critical field $H_{c2}$, determined principally by its penetration depth and coherence length, respectively. The superheating critical field of $MgB_2$, calculated from $H_{sh} = 0.75\sqrt{H_{c1}H_{c2}}$ [112, 113], is 170–1000 mT depending on the field's direction [114], which suggests that it might be possible to achieve 200 MV/m gradient [115]. $MgB_2$ is also an attractive choice for multilayer film coatings to benefit from the lower surface resistance, which was proposed by Gurevich [116].

The SRF properties of $MgB_2$ films formed by two different processes are being characterized [117]. X. X. Xi *et al.* of Temple University are developing the growth of $MgB_2$ via hybrid physical-chemical vapor deposition (HPCVD) [118]. Superconductor Technologies, Inc., has deposited $MgB_2$ using a reactive evaporation (RE) technique at 550°C [119]. The field-dependent rf surface impedance of single crystal $MgB_2$ samples have recently been measured at 7.5 GHz [120]. The road to understanding and exploiting this material yet lies ahead.

## 5.  Research on Fundamentals

### 5.1.  *Non-linear losses*

The typical SRF performance of a cavity is expressed in terms of the $Q_0$ as a function of the RF field, either $E_{acc}$ on-axis or peak surface electric ($E_p$) or magnetic ($B_p$). RF tests of bulk Nb cavities at or below 2.0 K show several non-linearities, referred to as "Q-slopes", occurring typically at low ($B_p < $~30 mT), medium (~30 $< B_p$ < ~90 mT) and high field ($B_p > $~90 mT) whose origins are not well understood. Figure 5 shows a typical $Q_0(B_p)$ plot for a bulk Nb cavity. Understanding the origin of these non-linearities is important not only from a scientific point of view but also to help identify better treatments to sustain high Q₀-values at high RF fields.

Low-β cavities and elliptical-type cavities tested at 4.2 K commonly show a monotonic decrease of $Q_0$ by about a factor of ~3-10 up to ~80-90 mT. This effect is attributed mainly to a thermal feedback in which $R_s(B_p)$ increases because of the warming of the inner cavity surface with increasing RF field due to the exponentially increasing BCS-surface resistance and the poor heat conductivity of He I [121].

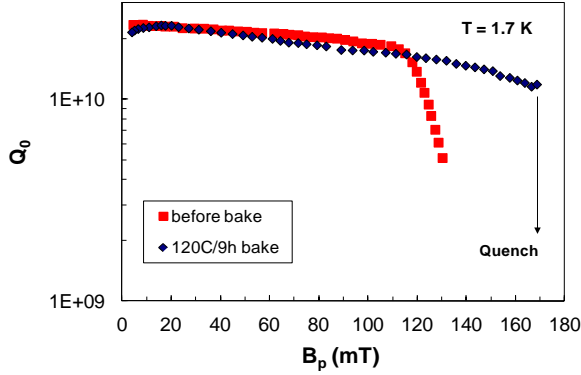

Fig. 5. Typical $Q_0(B_p)$ curves for a 1.5 GHz bulk Nb cavity before and after LTB (adapted from [122]).

Thin-film Nb magnetron-sputtered cavities exhibit a monotonic decrease of $Q_0$ by about one order of magnitude up to ~90 mT at or below 2.0 K and such strong degradation of $Q_0$ has precluded their application to accelerators which require $E_{acc}$ > ~15 MV/m [123]. The $R_s(E_{acc})$ plot for a state-of-the art Nb/Cu thin film cavity is shown in Fig. 6.

In spite of significant R&D efforts, mainly at CERN, there is no clear explanation for the origin of such losses. Possible hypothesis include suppression of the lower critical field, $H_{c1}$, and of the critical superfluid velocity, because of the low mean free path of the normal electrons in the films, or additional losses due to "weak-links" because of the granularity of the films [123], and fine-scale surface roughness. A recent analysis of field-dependent rf losses measured on magnetron sputtered Nb/Cu thin film samples at CERN [124] showed good agreement with a model based on electric field driven losses due to quasi-particle tunneling into localized states in the surface oxide (referred to as "interface tunnel exchange" model discussed in [125] and references therein).

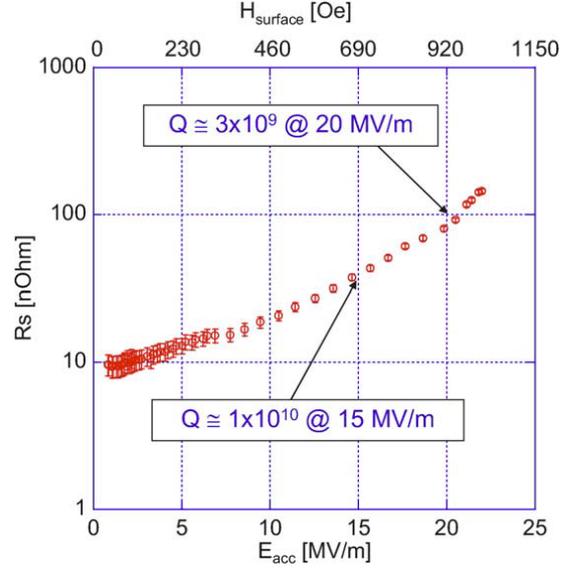

Fig. 6. State-of-the-art performance of 1.5 GHz Nb/Cu thin film cavity at 1.7 K. The figure was taken from [123].

### 5.1.1. *Low-field Q-slope*

An increase of the $Q_0$-value by up to ~50% between ~2 mT and ~20 mT is sometimes observed. This effect seems to be enhanced at lower temperatures and by a low-temperature (120-160°C) baking in ultra-high vacuum or atmospheric noble gas [86] for several hours (LTB).

In a model discussed in Ref. [125], this effect is related to non-equilibrium superconductivity caused by sub-oxides precipitates near the Nb surface. An alternative explanation, proposed in Ref. [88], relates it to the de-trapping of vortices which are within a depth of the order of the RF penetration depth from the surface, due to the Lorentz force of the RF field. According to this model, the $Q_0(B_p)$ curve should be hysteretic at low field, with a constant, high $Q_0$-value obtained after reducing the RF power back to low field.

An attempt at a systematic investigation of this phenomenon was published in Ref.[126], but further dedicated studies are necessary for a better understanding.

### 5.1.2. *Medium-field Q-slope*

A reduction of $Q_0$ by about 20-50% between ~20 mT and ~90 mT is typically observed. This phenomenon seems to be enhanced by EP and LTB and at lower temperatures while it is reduced in large-grain cavities. The $R_s$ field dependence is often well described as being quadratic, although a linear term was sometimes included, particularly to fit data after LTB [43]. The linear term was related to losses caused by Josephson fluxons at "strong" superconducting links near the Nb surface, consisting of oxide-filled grain boundaries [127].

The $R_s(B_p)$ dependence predicted by the thermal feedback model is not as strong as is often measured experimentally, particularly for cavities at gigahertz frequency for which $R_{BCS}$ is relatively low compared to S-band cavities [121]. The convenient availability of large Nb crystals has enabled a more precise characterization of the Kapitza resistance at the niobium/superfluid helium interface [128].

An intrinsic non-linearity of $R_{BCS}$ due to the increase in thermally-activated quasi-particles by the RF field was calculated in the clean limit in [129]. The model provides a good fit of the experimental data in the gigahertz range and predicts a stronger slope a lower temperatures, as it had been measured.

Although some common trends in the medium field Q-slope for different surface treatments and materials have been identified experimentally, a good understanding of which parameters are affected by such treatments/materials, in relation to the medium field Q-slope, is still missing.

### 5.1.3. *High-field Q-slope*

A sharp, exponential drop of $Q_0$ starting at ~90 mT in absence of field emission characterizes the performance of bulk Nb cavities tested after typical active chemical processing. Temperature mapping of the outer cavity surface during RF tests show non-uniform losses ("hot-spots") occurring in the high-magnetic field region of the cavities' surface. Since this phenomenon was first reported in 1997, many models have been proposed to explain the origin of the anomalous losses. Those includes defective oxides, high interstitial oxygen near the surface, reduced field of first flux penetration due to lattice defects or impurities near the surface, local quenches at sharp surface features (such as steps at grain boundaries). Extensive reviews of experimental data and models related to the high-field Q-slope have been presented over the years in many publications [87, 130-132]. The most recent model proposes a field dependent expression for the surface resistance which can describe the Q-slopes in any field region. At the origin of $R_s(B_p,T)$ are small (compared to the RF penetration depth and coherence length) isolated "defects" which suppress the surface barrier for flux penetration. $R_s$ is factorized in a temperature dependent term, resulting from the increase in the density of normal-conducting electrons above a certain percolation temperature, and a field dependent term resulting from the growth of the normal conducting volume with the applied RF magnetic field [133].

While some of the models provide a good fit of the experimental data in the high-field *Q*-slope region none of them has clearly provided a physical explanation of the phenomenon which is widely accepted, although hydrogen vacancy complexes in the near surface have been implicated [134].

It was found empirically that electropolishing plus LTB significantly reduces the high-field *Q*-slope of fine-grained Nb cavities (ingot Nb may not need EP), allowing attainment of $B_p$-values of up to ~190 mT at 2.0 K [87]. In addition to that, LTB causes a reduction of $R_{BCS}$ and often an increase of the residual resistance, $R_{res}$, so that a moderate 10-20% increase of $Q_0$ at 2.0 K, at low field, is obtained [88]. Lower $R_{res}$-values can be restored by rinsing the cavity with HF (49%) after baking [135]. A well accepted explanation of the baking effect is also missing. That the beneficial effect of LTB on the high-field Q-slope is found to be dependent on the material (fine-grain or ingot Nb) and treatment combination (EP, BCP or post-purification) [54] suggests that multiple mechanisms are involved in what is generally described as "high-field *Q*-slope."

### 5.1.4. *Non-linear BCS losses*

Xiao has recently derived a field-dependent extension of Mattis-Bardeen theory of a BCS superconductor's surface impedance [136]. Prior description of the response of a superconducting surface to RF fields derived from BCS theory has been in the low-field limit. Xiao finds as a result of the modified density of states distribution with flowing Cooper pairs, that the surface resistance of such a superconductor initially decreases with increasing field before increasing at further higher fields. Such behavior has not been previously predicted. An intriguing correspondence of this prediction with the frequently observed "low-field Q-slope" is noted. Analysis will continue.

## 5.2. *RF Critical Field*

Theoretically, the amplitude of the maximum RF magnetic field which can be applied at the surface of a type-II superconductor before the surface barrier to flux penetration vanishes is the so-called superheating field, $H_s$. The ratio of $H_s$ divided by the thermodynamic critical field, $H_c$ has been calculated near $T_c$ using the Ginzburg-Landau (GL) theory in the limits of small and large GL parameter, $\kappa_{GL}$ [137]. The calculation of $H_s$ at low temperature is quite complex, and it has been recently done in the clean limit [138] and as a function of impurities [139] in the limit of large $\kappa_{GL}$. The results show that:

$$H_s = 0.84 H_c \quad T \rightarrow 0, \text{ clean limit,} \quad (2)$$

whereas $H_s$ changes by less than ~3% in the dirty limit, in the presence of nonmagnetic impurities. Magnetic impurities, on the other hand, significantly suppress $H_s$. An interesting result of the calculation was that, unlike in the clean limit, the presence of nonmagnetic impurities maintains an energy gap in the quasi-particle spectrum near $H_s$, which would result in reduced $R_s$ non-linearity. To the authors' knowledge there exists no calculation of $H_s(T \ll T_c)$ as a function of $\kappa_{GL}$ in either clean or dirty limit.

The temperature dependence of $H_s$ can be approximated as:

$$H_s(T) \approx H_s(0)\left[1 - \left(\frac{T}{T_c}\right)^2\right]. \quad (3)$$

Experimentally, the superheating field has been achieved in SnIn and InBi alloy samples with different $\kappa_{GL}$ values and type I superconductors such as Sn, In and Pb near $T_c$ [140]. However, measurements of $H_s$ in Nb and Nb$_3$Sn cavities as a function of temperature showed $H_s$-values lower than predicted by the theory for $T \ll T_c$. The cause for this discrepancy is not clear [141].

The highest $B_p$-value measured on bulk Nb cavities at 2.0 K is 210 mT [142], greater than $H_c(2.0 \text{ K}) = 190$ mT.

While some measurements confirm the possibility of reaching the superheating field, as predicted theoretically, $H_s$ might still be of limited practical significance for SRF cavities: with a typical surface area of several meters square, it is easy to imagine locations with reduced surface barrier because of, for example, roughness or clusters of impurities. Furthermore, operating cavities close to $H_s$ will be impractical as $R_s$ approaches the normal-state surface resistance. Nevertheless, understanding $H_s$ both from the experimental and theoretical point of view is important, particularly for materials alternate to Nb with high $\kappa_{GL}$, which have $H_c$-values a factor of ~2 greater than Nb.

## 5.3. *Surface topography*

Surface treatments on bulk Nb cavities and substrate and deposition methods in thin-film cavities determine the topography of the outermost ~100 nm layer which carries the RF current. A comprehensive way to look at surface roughness over different length scales ("macro-roughness" vs. "micro-roughness") is to plot the power spectral density (PSD) as a function of the spatial frequency. The PSD is calculated from the surface profiles measured with different instruments, such as atomic force microscope and stylus profilometer, over different length scales. PSD data can be compared with different surface structure models to understand how processes determine the surface topography [143].

Data analysis clearly indicates that BCP, unlike EP, produces surface structures dominated by step edges, because of differential grain boundary etching. Smoothing by EP occurs mostly in the spatial frequency range corresponding to 1-10 μm [66]. Removal of only ~15 μm by EP is already sufficient to smooth sharp edges in BCP-treated samples, whereas the overall root-mean-square (RMS) roughness decreases steadily with increasing material removal by EP, up to ~50 μm [144].

The region of the cavity with the largest RMS step height is the equatorial weld, which is also the region where the surface magnetic field is close to the peak value. Samples' measurements show that ~90 μm of material removal by EP is required to reduce the RMS step height in the weld region to a value comparable to areas not affected by the welding [145].

Replica techniques have been developed by several laboratories to extract the topography of outstanding features on the inner surface of cavities which are often associated with quenches [146]. The resolution of the technique is ~1 μm and typical defects are "pits", ~100 μm in diameter, ~100 μm deep, usually found in the EBW heat-affected zone, as mentioned in Sec. 4.3

Systematic studies both at DESY [147] and KEK [42] show that the quench field of a cavity treated by EP and LTB decreases with increasing material removal by BCP, followed by LTB. This trend is inverted if additional material removal by EP is applied [148]. While these studies help making the case why "smoother is better", there exist examples of cavities reaching $E_{acc}$~40 MV/m having a "rough-looking" surface [149, 150]. Furthermore, no significant dependence in the performance of multi-cell ingot Nb cavities at DESY was found whether grain boundary steps were mechanically polished prior to BCP [30]. What remains yet elusive is clear discrimination of what scale roughness "matters." Efforts have begun to model specific non-linear losses due to specific representative topographies [151, 152].

Finally, it should be mentioned that changes in non-linear losses and the quench field associated with EP or BCP might also be related not only to topography but also to different concentration of interstitial impurities, such as hydrogen and oxygen, or lattice defects near the surface introduced by the two processes, but specific evidence is lacking.

## 5.4. *Grain boundaries*

The influence of grain boundaries on cavity performance is another controversial topic under investigation. Measurements on three single-cell cavities of the same shape made of fine-grain, large-grain and single-crystal Nb and treated in the same way did not show significant differences of performance [153].

As mentioned in the previous paragraph, sharp steps at grain boundaries cause a geometric local magnetic field enhancement which can lead to a premature quench, if the direction of the RF field is nearly normal to the grain boundary plane [154].

Magneto-optical [155] and flux-flow [156] studies on Nb samples showed that:

- Grain boundary steps of height greater than ~10 μm found on the fine-grain welded samples can induce preferential flux penetration in those regions.
- Preferential flux penetration at a grain boundary was clearly observed in bi-crystal samples when the grain boundary plane is close to parallel with the applied magnetic field. In this configuration, the dissipation due to flux-flow is the highest.

There exist no direct measurements of the grain boundary depairing current density, $J_b$, in bulk Nb. However, it can be estimated from the following formula, assuming that the grain boundary makes a superconductor-insulator-superconductor junction:

$$J_b = \frac{\pi \Delta}{2 e \, G_{gb}}, \qquad (4)$$

where $\Delta$ is the energy gap at 0 K, $e$ is the electron's charge and $G_{gb}$ is the grain boundary specific resistance. Taking $G_{gb} \cong 2 \times 10^{-13} \ \Omega \ m^2$ [157] and $\Delta = 1.55$ meV [158], one obtains $J_b \cong 1.2 \times 10^{10}$ A/m$^2$.

The critical depinning current density, $J_c$, obtained from magnetization measurements of cavity-grade Nb samples was at least a factor of ten lower than $J_b$ [29]. This would suggest that losses due to oscillations of pinned vortices under an RF field might be more significant for a material such as bulk Nb, compared to high-temperature superconductors with stronger pinning.

As mentioned in Sec. 4.2.1, the efficiency of trapping residual magnetic field when cooling Nb samples below $T_c$ was found to be lower in large-grain than fine-grain Nb. This effect, combined with the reduced grain boundary resistance contribution, because of fewer boundaries, might explain the lower $R_{res}$ often measured for ingot Nb cavities, compared to that of fine-grain cavities. In addition, the onset of the high-field Q-slope is typically higher in ingot Nb than fine-grain cavities, for the same surface treatment [43].

Oxide-filled, small size (<0.5 μm) grain boundaries in magnetron-sputtered thin-film Nb/Cu cavities have been described as creating "weak-links" which are sources of additional RF losses because of [159]:

- Grain boundary resistance contributing to the residual resistance at low field.
- Josephson or Abrikosov-Josephson fluxons penetrating along grain boundaries already at low field.

## 5.5. *Surface interstitials and oxide*

For typically prepared Nb surfaces the concentration near the surface of interstitial impurities, such as oxygen and hydrogen, is significantly higher than in the bulk of the material, even for high-RRR Nb.

Hydrogen is very mobile and segregates at the oxide/metal interface because of the local strain induced by the oxide layer and the presence of other impurities such as oxygen. It is very challenging to accurately measure hydrogen depth profiles in Nb, and results obtained by nuclear reaction analysis (NRA) show concentrations of ~40 at.% in a ~5 nm layer [126, 160]. Such concentration can lead to metallic hydride formation in the temperature range 90-150 K. NRA measurements also show that the hydrogen peak near the surface is reduced by a factor of ~3 by the LTB.

Interstitial oxygen concentrations of up to 2–10 at.% in a 1–10 nm layer below the oxide was found as a result of the growth of the surface oxide layer [161, 162]. High interstitial oxygen in Nb suppresses $T_c$ and $H_c$. As the oxygen diffusion length in Nb baked at 120 °C for 48 h is ~40 nm, it seemed plausible that oxygen diffusion might explain the baking effect mentioned in Sec. 5.1.3. However, measurements by time-of-flight secondary ion mass spectrometry (TOF-SIMS) [163] and X-ray scattering [164] do not support this hypothesis.

The quality of the first ~20 nm of Nb metal, and therefore of the metal/oxide interface seems to be crucial for the cavity performance. The value of the energy gap obtained from fits of $R_s(T)$ at low RF field with BCS theory numerical calculations clearly showed a correlation between high gap values and high maximum $B_p$-values [122]. Surface treatments such as BCP, EP, anodization, LTB do affect Δ and therefore the ultimate field in the cavity. Evidence of the influence of surface treatments on the quasi-particles density of states (DOS) has also been obtained by point-contact tunneling [158].

The niobium oxide layer has been studied to great extent since the 1970s. X-ray photoelectron spectroscopy (XPS) techniques (laboratory XPS, synchrotron XPS, angle-resolved XPS), transmission electron microscopy (TEM), 3D atom probe tomography (APT) have all been applied to study the oxide layer. The results indicate the presence of an amorphous niobium pentoxide layer, ~2-10 nm thick depending on the oxidation conditions, covering the Nb metal [165, 166]. The presence of "defects" in this layer, such as oxygen vacancies, was inferred from tunneling measurements.

Some studies show that the transition from $Nb_2O_5$ to Nb is fairly sharp, with no intermediate suboxide layers [167], while others suggest the presence of an intermediate ~1 nm $NbO_x$ layer [162]. Baking in UHV at 145 °C causes a partial decomposition of $Nb_2O_5$ into $NbO_2$ [168]. The oxygen concentration within 10 nm below the oxide was found to increase by no more than ~1.5 at.% [169]. Subsequent air exposure causes $Nb_2O_5$ to grow back to its initial thickness [170].

A systematic study to evaluate the influence of the oxide layer on $R_{res}$ determined that its contribution is < ~1.5 $n\Omega$ [171]. Losses due to interface tunnel exchange caused by defects in $Nb_2O_5$ had been proposed to explain the high-field Q-slope [125], but some cavity test results do not support this model [87, 130].

Recent tunneling spectroscopy measurements on chemically etched Nb samples showed the presence of quasi-particle states within the energy gap and a broadening of the BCS DOS peaks and are associated with surface paramagnetism [158, 172]. These results were explained by the presence of magnetic impurities near the surface which were attributed to substoichiometric $Nb_2O_5$. The results also showed that LTB significantly reduces those features in the DOS spectrum.

The metal/oxide interface and impurities in this region seem to have a significant effect on SRF properties of Nb. The above mentioned results suggest the need for treatments which reduce the hydrogen concentration, which is achieved to some degree by heat treatments, and allow the formation of less defective oxide, such as during a controlled "dry" oxidation.

### 5.6. *Magnetic screening by thin SC layers*

Efforts are underway to demonstrate and ultimately exploit the theoretical possibility of supporting much higher surface rf magnetic fields through the use of thin multi-layer superconductor/insulator structures as proposed by Gurevich [116]. Early measurements of the DC magnetic field properties of thin SC layers appear to be consistent with the theory, with inhibited flux entry into the surface [173, 174]. Realizing such hybrid structures will require incorporation of a host of techniques from, for example, the optical coating community. Key challenges will be securing consistent material properties of layers thinner than the magnetic penetration depth, $\lambda$, and sharp, smooth boundaries between layers. Work at ANL and Jefferson Lab seeks to produce candidate multilayer surfaces for rf testing [109, 175, 176].

## 6. Cavity Diagnostics

Several diagnostic methods have been developed or adapted to identify quench locations inside multi-cell cavities. Thermometry on the cavity exterior surface is still the most powerful technique to identify lossy regions of the cavity interior surface. A two-cell temperature mapping system consisting of 320 carbon resistor-temperature-devices (RTD) was developed at Jefferson Lab to monitor the temperature in the equator region of two out of nine cells in a 1.3 GHz cavity [177]. By measuring $Q_0(B_p)$ in all modes of the $TM_{010}$ passband, the quench location can be typically isolated to one of two cells. Two arrays with eight Cernox RTDs each have been used at Fermilab as a fast (thermal response time of ~15 ms) thermometry system to identify the quench location at the equator of a cell of a nine-cell 1.3 GHz cavity [58, 178]. A full nine-cell thermometry system with 4600 carbon RTDs has been built at Los Alamos [179]. A new flexible "octopus" thermometry array has recently been developed for use at CERN [180].

For quench location detection, oscillating superleak transducers (OST) were developed at Cornell as an alternative to RTDs. OSTs detect the He-II second sound wave driven by the defect-induced quench. By measuring the time of arrival of the second sound wave at three or more detectors the defect location can be unambiguously determined in three dimensions, with similar spatial resolution as by thermometry (~1 cm) [181].

Once the defect location has been located, imaging systems have been developed at Kyoto University and JLab to take pictures of the defect inside the cavity with a resolution lower than 10 $\mu$m [182, 183].

A new system developed by MicroDynamics Inc. and Jefferson Lab provides optical interferometric profilometry of the interior surface of multi-cell cavities. Dubbed CYCLOPS for CavitY CaLibrated Optical Profilometry System, the unit enables normal incident viewing and submicron surface topographic mapping with <2 µm lateral resolution. The system is intended both for detailed defect characterization and quantitative surface topography characterization as a function of applied surface treatment processes.

A local-grinding tool was also developed at KEK and was successfully used to repair a defect identified by thermometry, improving the performance of a 1.3 GHz nine-cell from 16 to 27 MV/m [184].

An X-ray diagnostic system is in routine use at KEK and is currently being developed at Jefferson Lab to identify the location of field emitters. By measuring the endpoint energy of x-ray spectra for all pass-band modes, one can localize the electron emission source to specific annular rings on the surface of multicell cavities. Further azimuthal resolution is under development.

At Jefferson Lab, a low-temperature laser scanning microscopy technique was recently applied to obtain maps of the surface resistance directly on the inner surface of a single-cell RF cavity [185].

Special rf structures have been developed which permit characterization of the SRF performance properties of small material samples. The motivation is to reduce the ambiguity and cost that comes with dependence on rf measurements with extended 3D structures such as are used for accelerators. These systems enable detailed studies of temperature, field, and frequency dependent properties of sample SRF materials and surfaces and allow direct correlation of performance with material characteristics discerned via surface analytic techniques [186-190].

Clearly, the evolution of new and improved diagnostic and repair tools enable systematic characterization of performance-limiting effects, which is the first step to improving both the performance envelope and the reliability of processes which produce SRF cavities.

## 7. Linkage of technical SRF performance with viable RF structure design options

One of the variously stimulating or vexing aspects of SRF science and technology for accelerators is the complexity of the multi-parameter design and performance space. Progress in understanding or control in one dimension creates new opportunities for useful solutions with other parameter changes.

For example, before surface cleaning techniques evolved adequately, multipacting and parasitic field emission drove structure design constraints. As process confidence grows, higher surface electric fields are "allowed" in SRF accelerating structure design which in turn extends the accessible structure parameter space, allowing for higher shunt impedance and design for minimum dissipative losses. Add to this improved fabrication processes and surface treatment processes and the door is opened to increased geometrical complexity of accelerator structures beyond the simple axisymmetric. Thus one begins to see performance previously associated with $\beta =1$ elliptical structures occurring in QWR, HWR, and spoke-type structures.

The ultimate performance of an SRF acceleration system is simply as a highly efficient transformer of supplied rf power into beam power. For some applications, even that is insufficient because the beam is itself but part of the transformer with the intended product being tuned photons, thus the interest in realizing ERLs. A very practical constraint on ERL power efficiency, however, is the usable loaded-$Q$ ($Q_l$) = $f$/bandwidth of the SRF cavity. With very light net beamloading, efficiency interests push for higher $Q_l$, but this must not come at the expense of increased detuning from drift or phase noise due to microphonics. For RF power efficiency, frequency stability needs to remain a small fraction of the resonance bandwidth. Thus the interest by Liepe *et al*. at Cornell to establish new design criteria for structure design and the placement of stiffening rings on elliptical structures for ERLs [191].

When techniques are established that yield the best bulk Nb performance from deposited conformal films on inexpensive high thermal conductivity substrates, the cost equation of SRF for accelerators will change dramatically. Cryomodule designs could shift to lower cost materials and greatly reduced touch labor. Such is not available today, but no fundamental physical properties stand in the way. Nevertheless, the realization of high-performing SRF thin film cavities may not be an "easy" task which will be accomplished rather quickly.

SRF cavities also present higher order modes (HOMs) to the beam. The beam/HOM interactions must be managed to avoid disruption of the beam and excessive losses. While resonant structure design considerations seek to minimize excitation of HOMs in a particular application setting by both field design and extraction of mode energy via rf couplers for dissipation in external loads, fabrication tolerances must also be taken into account to assure that such geometry-dependent properties are realized in practice [192, 193]. Dedicated international workshops focus on these issues [194].

## 8.  Potential Applications of SRF Technology

As SRF technology matures, the variety of accessible accelerator applications grows. The technology becomes less and less the domain of self-contained research institutions and increasingly suitable for production systems and industrial operations. In the US, the broad topic of accelerator applications is being encouraged by the US Department of Energy [195]. Similar governmental support is expressed around the world because of the perceived economic potential. A large fraction of the systems under discussion require SRF for the simple reason of net energy efficiency. For realization, large scale service systems require technological robustness and unambiguous cost effectiveness—such appear within reach for SRF.

We identify here some representative applications of SRF that are soon to arrive as well as some which are in various stages of planning, analysis or consideration. With the exception of the ILC, most of these applications require "moderate" accelerating gradient but rely on high $Q_0$-values for most efficient operation.

### 8.1.  *QWR for high-β hadrons*

Innovative designs of two QWR for use in RHIC are under development. A 56 MHz QWR will be used to maintain stored beam. A 28 MHz folded quarter wave structure with a large tuning range to accommodate particle revolution frequency change in RHIC during beam acceleration is also under development in collaboration with Niowave Inc.

### 8.2.  *European Spallation Source (ESS)*

In 2003 the joint European effort to design a European Spallation Source resulted in a set of detailed design reports. Lund was agreed as the site in 2009. The current baseline for the design delivers 5 MW of 2.5 GeV protons to a single target, in 2.86 ms long pulses with a 14 Hz repetition rate. The linac will have a normal conducting front end with an ion source, an RFQ and a DTL. The superconducting part starts with spoke cavities followed by two families of elliptical cavities. It will be the first time that spoke cavities are used in a major accelerator. The target completion date is 2018 [196].

### 8.3.  *Energy-Recovering Linacs*

Energy-recovering linacs (ERLs) are very attractive and efficient means of providing extremely high brightness and/or ultra-short pulse length light sources. Several projects are underway around the world [197]. Cornell University, for example, is developing the superconducting RF technology required for the construction of a 100 mA hard X-ray light source driven by an Energy-Recovery Linac [198]. The 5 GeV ERL main linac will have 384 7-cell SRF cavities, running CW at 16.2 MV/m accelerating gradient. A compact ERL (cERL) is under construction at KEK as a demonstration vehicle for technologies needed for a potential 3 GeV ERL [199].

A solid conceptual design was developed for an ERL-based 4GLS multi-spectrum photon source to have been located at Daresbury [200].

## 8.4. *Crabbing and deflecting cavities*

In contrast to the majority of SRF accelerator applications, there are also major efforts to exploit high field dipole modes to give bunches a head/tail differential transverse kick in order to maximize luminosity of colliding beams with non-zero crossing angle. This so-called crabbing scheme has been demonstrated at KEK [201] and is under active development for the APS SPX upgrade [202] and the LHC luminosity upgrade [203].

There is also a range of applications of these cavities in deflecting mode such as for beam separation and bunch length measurement [204]. A recent analysis of compact deflecting and crabbing SRF cavities was contributed by Delayen [205].

## 8.5. *SRF photoinjectors*

In the pursuit of ever lower emittance electron sources, it is attractive to embed a photoelectron source directly in an SRF cavity as a means of providing very high fields right at the emission surface. The extreme vacuum of the SRF environment is also attractive for minimizing low-energy scattering effects. Arnold and Teichert have provided a recent overview of progress in this area [206]. SRF photoinjectors are being actively developed at Helmholtz-Zentrum-Berlin, Brookhaven National Laboratory, Helmholtz-Zentrum Dresden-Rossendorf, PKU, and University of Wisconsin [207-210].

## 8.6. *NGLS*

LBNL is developing design concepts for a multi-beamline soft x-ray FEL array powered by a CW superconducting linear accelerator, operating with a high bunch repetition rate of approximately one MHz [211]. A nominal electron beam energy of 2.4 GeV has been chosen.

## 8.7. *Upgrade of LANSCE at LANL*

By increasing LANSCE H+/H- linac beam power from 800 kW to 2 MW using SRF cavities, LANL will be able to achieve neutron flux and irradiation volume equivalents similar to those expected to be achieved by the IFMIF (International Fusion Materials Irradiation Facility) [212]. By increasing LANSCE proton energy from 800 MeV to 3 GeV using SRF cavities, higher resolution proton radiography can be achieved [213].

## 8.8. *Compact 4K light sources*

The principal attractive features of SRF accelerators tend to be efficient conversion of RF to beam power and high fields for compact footprint. As the materials and techniques evolve toward higher SRF performance at higher temperatures, new compact CW accelerator applications become viable. A compact X-ray source based on inverse Compton scattering of a high-power laser on a high-brightness linac beam has been proposed by MIT. The key enabling technologies are a high average power laser and a superconducting accelerator [214, 215].

## 8.9. *Photonic bandgap*

Employing a novel concept, 2.1 GHz superconducting RF photonic band gap (SRF PBG) structures have been designed, fabricated and tested by LANL and Niowave Inc. The maximum $E_{acc}$ of 15 MV/m has been achieved with predicted $Q_0$. This structure can reduce long-range wakefields and increase current threshold for high-current SRF accelerators [216, 217] .

## 8.10. *ADS*

Accelerator driven systems (ADS) are increasingly considered to be promising means for the transmutation of nuclear waste, and also as credible schemes for Th-based energy production. Technical preparation work for ADS has been taken up by the international accelerator community. SRF proton linacs are almost generally considered as the right answer to the ADS challenges: very high availability multi-MW CW proton beams, with energy around 1 GeV. The high modularity of such linacs is a key issue [218]. Significant efforts are underway in India, China, and Europe [219-221].

In the US, a technology readiness assessment for ADS systems was composed in 2010 [222].

### 8.11.  *Project X*

A proposed Project X at FNAL will require a 3 GeV, 1 mA CW proton LINAC based on SRF technology. It will utilize a wide range of cavity types, including 162.5 MHz HWRs at $\beta$=0.1, 325 MHz single spoke resonators at $\beta$=0.22 and $\beta$=0.47 and 650 MHz elliptical cavities at $\beta$=0.61 and $\beta$=0.9. The project could be ready for a construction start as early as 2016. Possible extension to a future pulsed LINAC to increase the beam energy up to 8 GeV would require 1.3 GHz $\beta$=1 elliptical cavities [223]. Design and specifications of the cavities for this project are driven by the large dynamic heat loads, emphasizing the need to achieve high $Q_0$-values.

### 8.12.  *CERN-based possibilities*

Several large future facilities at CERN would require SRF-based systems. LHeC, a proposed e-p collider based on the LHC plus a new superconducting e- accelerator, based either on a ring or an ERL are in planning [224]. SPL, a possible new injector chain and high intensity proton source remains on the candidate list [225]. Discussions have begun regarding a LEP3 $e^+e^-$ collider targeted specifically for Higgs production. Such would require new SRF systems [226].

### 8.13.  *ILC*

The largest single application of SRF technology being discussed is the proposed International Linear Collider (ILC). In 2004 ICFA chartered the International Linear Collider Steering Committee, which established the Global Design Effort (GDE) to develop a technical design for a 0.5 TeV $e^+e^-$ collider. The focused international collaborative effort built on the former TESLA program produced significant R&D progress and will produce a Technical Design Report at the end of 2012 [227, 228]. With the TDR in hand, the community awaits LHC physics results to sharpen the motivation and target for an ILC. Meanwhile, sustaining efforts will pursue cost-saving technology improvements.

### 8.14.  *Muon collider*

Lepton colliders are the cleanest way to probe the energy frontier. With mass 207 relative to that of electrons, muons suffer negligible synchrotron radiation loss. A multi-TeV $\mu^+\mu^-$ collider could be circular and therefore have a relatively compact geometry and employ multi-pass acceleration, reducing the infrastructure cost for attaining TeV energies. The critical challenge of such a scheme is 6-dimensional phase space cooling of the muons sufficient to create an effective beam. Large acceptance (~200 MHz) SRF cavities would be required. Costs for such motivate pursuit of Nb/Cu and other thin film SRF solutions [229-231].

## 9.  Conclusion

As understanding of key materials and processes grow, the community's ability to define and design methods which yield tailored SRF-based accelerator systems that support high amplitude and well-shaped rf fields with increasingly efficient power consumption also matures and flowers into a remarkable variety of useful solutions. We have attempted to offer a brief review of the current state of SRF R&D in support of future accelerators. There is, no doubt, additional high-quality work that we have not mentioned. The single best written record of the evolution of SRF-related science and technology is captured in the proceedings of the SRF Workshop/Conference series, now archived together with the LINAC and (x)PAC conferences on the JACoW website [232].

## Acknowledgments

Numerous colleagues have provided assistance identifying the latest references available in this fast-developing technology, we are grateful for their assistance assembling this review. This paper is authored by Jefferson Science Associates, LLC under U.S. DOE Contract No. DE-AC05-06OR23177. The U.S. Government retains a non-exclusive, paid-up, irrevocable, world-wide license to publish or reproduce this manuscript for U.S. Government purposes.

## References

[1] *The Technical Design Report of the European XFEL* (DESY XFEL Project Group, 2007) DESY 2006-097, http://xfel.desy.de/technical_information/tdr/tdr/.

[2] D. Reschke, Challenges in SRF Module Production for the European XFEL, *Proc. 15th Int. Conf. on RF Superconductivity,* Chicago, IL USA (2011), pp. 2-5, http://accelconf.web.cern.ch/AccelConf/SRF2011/papers/moioa01.pdf.

[3] C. E. Reece, SRF Challenges for Improving Operational Electron Linacs, *Proc. 15th Int. Conf. on RF Superconductivity,* Chicago, IL USA (2011), pp. 14-19, http://accelconf.web.cern.ch/AccelConf/SRF2011/papers/moioa04.pdf.

[4] J. Hogan, A. Burrill, K. Davis, M. Drury, and M. Wiseman, 12 GeV Upgrade Project - Cryomodule Production, *Proc. Int. Part. Accel. Conf. 2012,* New Orleans, Louisiana (2012), pp. 2429-2431, http://accelconf.web.cern.ch/AccelConf/IPAC2012/papers/weppc092.pdf.

[5] C. Hovater, et al., Commissioning and Operation of the CEBAF 100 MV Cryomodules, *Proc. Int. Part. Accel. Conf. 2012,* New Orleans, Louisiana (2012), pp. 2432-2434, http://accelconf.web.cern.ch/AccelConf/IPAC2012/papers/weppc093.pdf.

[6] D. Longuevergne, C. D. Beard, A. Grassellino, P. Kolb, R. E. Laxdal, and V. Zvyagintsev, RF Cavity Performance in the ISAC-II Superconducting Heavy Ion Linac, *Proc. XXV Linear Accel. Conf.,* Tsukuba, Japan (2010), pp. 860-862, http://accelconf.web.cern.ch/AccelConf/LINAC2010/papers/thp044.pdf.

[7] G. Bisoffi, G. Bassato, S. Canella, D. Carlucci, A. Facco, P. Modanese, A. Pisent, A. M. Porcellato, and P. A. Posocco, ALPI QWR and S-RFQ Operating Experience, *Proc. 13th Int. Workshop on RF Superconductivity,* Peking Univ., Beijing, China (2007), pp. 55-62, http://accelconf.web.cern.ch/AccelConf/srf2007/PAPERS/MO404.pdf.

[8] P. Baudrenghien and e. al., The LHC RF System - Experience with Beam Operation, *Proc. 2nd Int. Part. Accel. Conf.,* San Sebastián (2011), pp. 202-204, http://accelconf.web.cern.ch/AccelConf/IPAC2011/papers/mopc054.pdf.

[9] P. Maesen, E. Ciapala, and G. Pechaud, Final tests and commissioning of the 400MHz LHC superconducting cavities, *Proc. 13th Int. Workshop on RF Superconductivity,* Peking Univ., Beijing, China (2007), http://accelconf.web.cern.ch/AccelConf/srf2007/PAPERS/WEP25.pdf.

[10] S. A. Belomestnykh, et al., Commissioning and Operations Results of the Industry-Produced CESR-Type SRF Cryomodules, *Proc. 2005 Part. Accel. Conf.,* Knoxville, TN (2005), pp. 4233-4235, http://accelconf.web.cern.ch/AccelConf/p05/PAPERS/TPPT089.PDF.

[11] J. D. Fuerst, The ATLAS Energy Upgrade Cryomodule, *Proc. 14th Int. Conf. on RF Superconductivity,* Berlin, Germany (2009), pp. 52-56, http://accelconf.web.cern.ch/AccelConf/SRF2009/papers/moocau04.pdf.

[12] K. Honkavaara, B. Faatz, J. Feldhaus, S. Schreiber, R. Treusch, and M. Vogt, Status of FLASH, *Proc. Int. Part. Accel. Conf. 2012,* New Orleans, Louisiana (2012), pp. 1715-1717, http://accelconf.web.cern.ch/AccelConf/IPAC2012/papers/tuppp052.pdf.

[13] E. Vogel and e. al., Test and Commissioning of the Third Harmonic RF System for FLASH, *Proc. Int. Part. Accel. Conf. 2010,* New Orleans, Louisiana, USA (2010), pp. 4281-4283, http://accelconf.web.cern.ch/AccelConf/IPAC10/papers/thpd003.pdf.

[14] Y. Zhang, Experience and Lessons with the SNS Superconducting Linac *Proc. 1st Int. Part. Accel. Conf.,* Kyoto, Japan (2010), pp. 26-30, http://accelconf.web.cern.ch/AccelConf/IPAC10/papers/mozmh01.pdf.

[15] R. L. Geng, J. Dai, G. V. Eremeev, and A. D. Palczewski, Progress of ILC High Gradient SRF Cavity R&D at Jefferson Lab, *Proc. 2nd Int. Part. Accel. Conf.,* San Sebastián, Spain (2011), pp. 334-336, http://accelconf.web.cern.ch/AccelConf/IPAC2011/papers/mopc111.pdf.

[16] Y. Yamamoto, H. Hayano, E. Kako, S. Noguchi, T. Shishido, and K. Watanabe, Progress of High Gradient Performance in STF 9-cell Cavities at KEK, *Proc. Int. Part. Accel. Conf. 2012,* New Orleans, Louisiana (2012), pp. 2233-2235, http://accelconf.web.cern.ch/AccelConf/IPAC2012/papers/weppc013.pdf.

[17] J. P. Ozelis, Gradient R&D in the US, *Proc. 15th Int. Conf. on RF Superconductivity,* Chicago, IL (2011), pp. 849-853, http://accelconf.web.cern.ch/AccelConf/SRF2011/papers/thioa02.pdf.

[18] D. Reschke, et al., Results on Large Grain Nine-Cell Cavities at DESY: Gradients up to 45 MV/m after Electropolishing, *Proc. 15th Int. Conf. on RF Superconductivity,* Chicago, IL USA (2011), pp.

490-494,
http://accelconf.web.cern.ch/AccelConf/SRF2011/papers/tupo046.pdf.

[19] M. P. Kelly, Z. A. Conway, S. M. Gerbick, M. Kedzie, R. C. Murphy, B. Mustapha, P. N. Ostroumov, and T. Reid, SRF Advances for ATLAS and other β<1 Applications, *Proc. 15th Int. Conf. on RF Superconductivity,* Chicago, IL USA (2011), pp. 680-683, http://accelconf.web.cern.ch/AccelConf/SRF2011/papers/thiob04.pdf.

[20] A. Facco, et al., Superconducting Resonators Development for the FRIB and ReA Linacs at MSU: Recent Achievements and Future Goals, *Proc. Int. Part. Accel. Conf. 2012,* New Orleans, Louisiana (2012), pp. 61-63, http://accelconf.web.cern.ch/AccelConf/IPAC2012/papers/mooac03.pdf.

[21] T. Khabiboulline, et al., High Gradient Tests of the Fermilab SSR1 Cavity, *Proc. Intl. Particle Accel. Conf. 2012,* New Orleans, Louisiana (2012), pp. 2330-2332, http://accelconf.web.cern.ch/AccelConf/IPAC2012/papers/weppc052.pdf.

[22] W. Singer, A. Brinkmann, D. Proch, and X. Singer, *Physica C* **386**, 379 (2003),

[23] H. Padamsee, J. Knobloch, and T. Hays, *RF Superconductivity for Accelerators* (J. Wiley & Sons, New York, 1998).

[24] C. C. Koch, J. O. Scarbrough, and D. M. Kroeger, *Phys. Rev. B* **9**, 888 (1974), http://link.aps.org/doi/10.1103/PhysRevB.9.888.

[25] J. Knobloch, The "Q disease" in Superconducting Niobium RF Cavities, Newport News, VA (2002), pp. 133-150, http://dx.doi.org/10.1063/1.1597364.

[26] R. Grill, W. Simader, M. Heilmaier, D. Janda, W. Singer, and X. Singer, Correlation of Microstructure, Chemical Composition and RRR-Value in High Purity Niobium (Nb-RRR), *Proc. 15th Int. Conf. on RF Superconductivity,* Chicago, IL USA (2011), pp. 863-867, http://accelconf.web.cern.ch/AccelConf/SRF2011/papers/thpo059.pdf.

[27] G. R. Myneni, P. Kneisel, and T. Carneiro, USA Patent No. 8128765B2 (2012).

[28] P. Kneisel, G. Ciovati, G. Myneni, T. Carneiro, and J. Sekutowicz, Preliminary Results from Single Crystal and Very Large Crystal Niobium Cavities, *Proc. 2005 Part. Accel. Conf.,* Knoxville (2005), pp. 3991-3993, http://accelconf.web.cern.ch/AccelConf/p05/PAPERS/TPPT076.PDF.

[29] S. Aderhold, Large Grain Cavities: Fabrication, RF Results and Optical Inspection, *Proc. 15th Int. Conf. on RF Superconductivity,* Chicago, IL USA (2011), pp. 607-610, http://accelconf.web.cern.ch/AccelConf/SRF2011/papers/weiob05.pdf.

[30] W. Singer, et al., Advances in large grain resonators for the European XFEL, *Proc. 1st Int. Symp. on the Supercond. Science and Technol. of Ingot Niobium,* Newport News, VA (2010), pp. 13-24, http://dx.doi.org/10.1063/1.3579220.

[31] G. Ciovati, P. Kneisel, and G. R. Myneni, America's overview of superconducting science and technology of ingot niobium, *Proc. 1st Int. Symp. on the Supercond. Science and Technol. of Ingot Niobium,* Newport News, VA (2010), pp. 25-37, http://dx.doi.org/10.1063/1.3579221.

[32] R. Geng, G. V. Eremeev, P. Kneisel, K. Liu, X. Lu, and K. Zhao, $Q_0$ improvement of large-grain multi-cell cavities by using JLab's standard ILC EP processing, *Proc. 15th Int. Conf. on RF Superconductivity,* Chicago, IL (2011), pp. 501-503, http://accelconf.web.cern.ch/AccelConf/SRF2011/papers/tupo049.pdf.

[33] P. Dhakal, G. Ciovati, and G. R. Myneni, A path to higher Q0 with large grain niobium cavities, *Proc. Int. Part. Accel. Conf. 2012,* New Orleans, Louisiana (2012), pp. 2426-2428, http://accelconf.web.cern.ch/AccelConf/IPAC2012/papers/weppc091.pdf.

[34] A. S. Dhavale, P. Dhakal, A. A. Polianskii, and G. Ciovati, *Supercond. Sci. Technol.* **25**, 065014 (10pp) (2012), http://iopscience.iop.org/0953-2048/25/6/065014.

[35] S. Aull, J. Knobloch, and O. Kugeler, Study of trapped magnetic flux in superconducting niobium samples, *Proc. 15th Int. Conf. on RF Superconductivity,* Chicago, IL (2011), pp. 702-706, http://accelconf.web.cern.ch/AccelConf/SRF2011/papers/thpo006.pdf.

[36] W. Singer, et al., Large grain superconducting RF cavities at DESY, *Proc. Int. Workshop on Single Crystal-Large Grain Niobium Technology,* Araxa, Brasil (2006), pp. 123-132, http://dx.doi.org/10.1063/1.2770685.

[37] S. K. Chandrasekaran, T. R. Bieler, C. Compton, W. Hartung, and N. T. Wright, Comparison of the role of moderate heat treatment temperatures on the thermal conductivity of ingot niobium, *Proc. 1st Int. Symp. on the Supercond. Science and Technol. of Ingot Niobium,* Newport News, VA (2010), pp. 131-141, http://dx.doi.org/10.1063/1.3579231.

[38] P. Dhakal, G. Ciovati, P. Kneisel, and G. R. Myneni, Superconducting DC and RF properties of ingot niobium, *Proc. 15th Int. Conf. on RF Superconductivity,* Chicago, IL (2011), pp. 856-861, http://accelconf.web.cern.ch/AccelConf/SRF2011/papers/thpo057.pdf.

[39] P. Kneisel, G. R. Myneni, G. Ciovati, J. Sekutowicz, and T. Carneiro, Development of

large grain/single crystal niobium cavity technology at Jefferson Lab, *Proc. Int. Workshop on Single Crystal-Large Grain Niobium Technology,* Araxa, Brasil (2006), pp. 84-97, http://dx.doi.org/10.1063/1.2770681.

[40] W. Singer, X. Singer, and P. Kneisel, A single crystal niobium RF cavity of the TESLA shape, *Proc. Int. Workshop on Single Crystal-Large Grain Niobium Technology,* Araxa, Brasil (2006), pp. 133-140, http://dx.doi.org/10.1063/1.2770686.

[41] P. Kneisel, G. Ciovati, W. Singer, X. Singer, D. Reschke, and A. Brinkmann, Performance of single crystal niobium cavities, *Proc. 2008 European Part. Accel. Conf.,* Genoa, Italy (2008), pp. 877-879, http://accelconf.web.cern.ch/accelconf/e08/papers/mopp136.pdf.

[42] H. Umezawa, K. Takeuchi, K. Saito, F. Furuta, T. Konomi, and K. Nishimura, Single crystal niobium development, *Proc. 1st Int. Part. Accel. Conf.,* Kyoto, Japan (2010), pp. 438-440, http://accelconf.web.cern.ch/accelconf/IPAC10/papers/mopeb073.pdf.

[43] H. Padamsee, *RF Superconductivity. Science, Technology, and Applications* (WILEY-VCH, Weinheim, 2009).

[44] E. Kako, S. Noguchi, M. Ono, K. Saito, T. Shishido, H. Safa, J. Knobloch, and L. Lilje, Improvement of cavity performance in the Saclay/Cornell/DESY's SC cavities, *Proc. 9th Int. Workshop on RF Superconductivity,* Santa Fe, NM (1999), pp. 179-186, http://accelconf.web.cern.ch/accelconf/SRF99/papers/tup011.pdf.

[45] W. Singer, D. Proch, and A. Brinkmann, Diagnostic of defects in high purity niobium, *Proc. 8th Int. Workshop on RF Superconductivity,* Abano Terme, Italy (1997), pp. 850-863, http://accelconf.web.cern.ch/accelconf/SRF97/papers/srf97d07.pdf.

[46] S. B. Roy, V. C. Sahni, and G. R. Myneni, Research & development on superconducting niobium materials via magnetic measurements, *Proc. 1st Int. Symp. on the Supercond. Science and Technol. of Ingot Niobium,* Newport News, VA (2010), pp. 56-68, http://dx.doi.org/10.1063/1.3579224.

[47] P. Kneisel, G. Ciovati, G. R. Myneni, W. Singer, X. Singer, D. Proch, and T. Carneiro, Influence of Ta content in high purity niobium on cavity performance, *Proc. 2005 Part. Accel. Conf.,* Knoxville, TN (2005), pp. 3955-3957, http://accelconf.web.cern.ch/accelconf/p05/PAPERS/TPPT075.PDF.

[48] T. R. Bieler, et al., *Phys. Rev. ST Accel. Beams* **13**, 031002 (2010), http://link.aps.org/doi/10.1103/PhysRevSTAB.13.031002.

[49] G. R. Myneni and H. Umezawa, *Materiaux & Techniques* **91**, 19 (2003),

[50] T. Gnaupel-Herold, G. R. Myneni, and R. E. Ricker, Investigation of residual stresses and mechanical properties of single crystal niobium for SRF cavities, Araxa, Brasil (2006), pp. 48-59, http://dx.doi.org/10.1063/1.2770678.

[51] High RRR niobium material studies, 2002,

[52] G. R. Myneni, Elasto-plastic behavior of high RRR niobium: effects of crystallographic texture, microstructure and hydrogen concentration, Newport News, VA (2002), pp. 227-239, http://dx.doi.org/10.1063/1.1597371.

[53] R. L. Geng and A. C. Crawford, Standard Procedures of ILC High Gradient Cavity Processing at Jefferson Lab, *Proc. 15th Int. Conf. on RF Superconductivity,* Chicago, IL USA (2011), pp. 391-393, http://accelconf.web.cern.ch/AccelConf/SRF2011/papers/tupo015.pdf.

[54] A. Romanenko and H. Padamsee, *Supercond. Sci. Technol.* **23**, 045008 (4pp) (2010), http://iopscience.iop.org/0953-2048/23/4/045008.

[55] X. Zhao, G. Ciovati, and T. R. Bieler, *Phys. Rev. ST Accel. Beams* **13**, 124701 (2010), http://prst-ab.aps.org/abstract/PRSTAB/v13/i12/e124701.

[56] L. D. Cooley, et al., *IEEE Trans. Appl. Superconductivity* **21**, 2609 (2011), http://dx.doi.org/10.1109/TASC.2010.2083629.

[57] R. L. Geng, et al., Latest results of ILC high-gradient R&D 9-cell cavities at JLAB, *Proc. 13th Int. Workshop on RF Superconductivity,* Beijing, China (2007), pp. 525-529, http://accelconf.web.cern.ch/accelconf/srf2007/PAPERS/WEP28.pdf.

[58] J. P. Ozelis, et al., Initial results from Fermilab's vertical test stand for SRF cavities, *Proc. 13th Int. Workshop on RF Superconductivity,* Beijing, China (2007), pp. 472-476, http://accelconf.web.cern.ch/accelconf/srf2007/PAPERS/WEP15.pdf.

[59] D. Kang, D. C. Baars, and C. C. Compton, Characterization of large grain Nb ingot microstructure using EBSP mapping and Laue camera methods, *Proc. 1st Int. Symp. on the Supercond. Science and Technol. of Ingot Niobium,* Newport News, VA (2010), pp. 90-99, http://dx.doi.org/10.1063/1.3579228.

[60] D. Kang, D. C. Baars, T. R. Bieler, G. Ciovati, C. Compton, T. L. Grimm, and A. A. Kolka, Characterization of Large Grain Nb Ingot Microstructure Using OIM and Laue Methods, *Proc. 15th Int. Conf. on RF Superconductivity,* Chicago, IL USA (2011), pp. 890-897, http://accelconf.web.cern.ch/AccelConf/SRF2011/papers/thpo067.pdf.

[61] A. Brinkmann, A. Gössel, W.-D. Möller, M. Pekeler, and D. Proch, Performance degradation in

several TESLA 9-cell cavities due to weld imperfections, *Proc. 8th Int. Workshop on RF Superconductivity,* Abano Terme, Italy (1997), pp. 452-456,
http://accelconf.web.cern.ch/accelconf/SRF97/papers/srf97c05.pdf.

[62]   W. Singer, *Physica C: Superconductivity* **441**, 89 (2006),
http://www.sciencedirect.com/science/article/pii/S0921453406001584.

[63]   V. Palmieri, Advancements on spinning of seamless multicell reentrant cavities, *Proc. 11th Int. Workshop on RF Superconductivity,* Travemunde, Germany (2003), pp. 357-361,
http://accelconf.web.cern.ch/accelconf/SRF2003/papers/tup26.pdf.

[64]   W. Singer, Progress in seamless RF cavities, in *13th International Workshop on RF Superconductivity,* edited by J. K. Hao, S. L. Huang and K. Zhao, Beijing, China, (2007), p. WE301,
http://accelconf.web.cern.ch/accelconf/srf2007/TALKS/WE301_TALK.pdf.

[65]   T. Higuchi, K. Saito, S. Noguchi, M. Ono, E. Kako, T. Shishido, Y. Funahashi, H. Inoue, and T. Suzuki, Investigation on Barrel Polishing for Superconducting Niobium Cavities *Proc. 7th Int. Workshop on RF Superconductivity,* Gif-sur-Yvette, France (1995), pp. 723,
http://accelconf.web.cern.ch/AccelConf/SRF95/papers/srf95f34.pdf.

[66]   F. Furuta, et al., Optimization of Surface Treatment of High-Gradient Single-Cell Superconducting Cavities at KEK *Proc. 2006 Linear Accel. Conf.,* Knoxville, Tennessee USA (2006), pp. 299,
http://accelconf.web.cern.ch/AccelConf/l06/PAPERS/TUP025.PDF.

[67]   F. Furuta, K. Saito, T. Saeki, H. Inoue, Y. Morozumi, Y. Higashi, and T. Higo, High Reliable Surface Treatment Recipe of High Gradient Single Cell SRF Cavities at KEK, *Proc. 13th Int. Workshop on RF Superconductivity,* Peking Univ., Beijing, China (2007), pp. 125-131,
http://accelconf.web.cern.ch/AccelConf/srf2007/PAPERS/TUP10.pdf.

[68]   A. D. Palczewski, H. Tian, and R. L. Geng, Optimizing Centrifugal Barrel Polishing for Mirror Finish SRF Cavity and RF Tests at Jefferson Lab, *Proc. Int. Part. Accel. Conf. 2012,* New Orleans, Louisiana (2012), pp. 2435-2437,
http://accelconf.web.cern.ch/AccelConf/IPAC2012/papers/weppc094.pdf.

[69]   *Mirror Smooth Superconducting RF Cavities by Mechanical Polishing with Minimal Acid Use* (FNAL, 2012) FERMILAB-PUB-11-032-TD,
http://lss.fnal.gov/archive/2011/pub/fermilab-pub-11-032-td.pdf.

[70]   R. Geng, B. Clemens, H. Hayano, K. Watanabe, and C. Cooper, Gradient Improvement by Removal of Identified Local Defects, *Proc. 15th Int. Conf. on RF Superconductivity,* Chicago, IL (2011), pp. 436-438,
http://accelconf.web.cern.ch/AccelConf/SRF2011/papers/tupo029.pdf.

[71]   l. Malloch, L. J. Dubbs, K. Elliott, R.Oweiss, and L.Popielarski, Niobium Reaction Kinetics: An Investigation into the Reactions Between Buffered Chemical Polish and Niobium and the Impact on SRF Cavity Etching, *Proc. Int. Part. Accel. Conf. 2012,* New Orleans, Louisiana (2012), pp. 2360-2362,
http://accelconf.web.cern.ch/AccelConf/IPAC2012/papers/weppc066.pdf.

[72]   K. Saito, et al., RD of Superconducting Cavities at KEK, *Proceedings of the Fourth Workshop on RF Superconductivity,* KEK, Tsukuba, Japan (1989), pp. 635-694,
http://accelconf.web.cern.ch/AccelConf/SRF89/papers/srf89g18.pdf.

[73]   H. Tian, S. G. Corcoran, C. E. Reece, and M. J. Kelley, *Journal of The Electrochemical Society* **155**, D563 (2008),
http://link.aip.org/link/?JES/155/D563/1.

[74]   H. Tian and C. E. Reece, *Phys. Rev. ST Accel. Beams* **13**, 083502 (2010),
http://link.aps.org/doi/10.1103/PhysRevSTAB.13.083502.

[75]   F. Eozénou, S. Berry, C. Antoine, Y. Gasser, J. P. Charrier, and B. Malki, *Phys. Rev. ST Accel. Beams* **13**, 083501 (2010),
http://link.aps.org/doi/10.1103/PhysRevSTAB.13.083501.

[76]   C. E. Reece and H. Tian, Exploiting New Electrochemical Understanding of Niobium Electropolishing for Improved Performance of SRF Cavities for CEBAF, *Proc. XXV Linear Accel. Conf.,* Tsukuba, Japan (2010), pp. 779-781,
http://accelconf.web.cern.ch/AccelConf/LINAC2010/papers/thp010.pdf.

[77]   H. Tian and C. E. Reece, Quantitative EP Studies and Results for SRF Nb Cavity Production, *Proc. 15th Int. Conf. on RF Superconductivity,* Chicago, IL USA (2011), pp. 565-570,
http://accelconf.web.cern.ch/AccelConf/SRF2011/papers/weioa01.pdf.

[78]   A. Reilly, T. Bass, A. Burrill, K. Davis, F. Marhauser, C. E. Reece, and M. Stirbet, Preparation and Testing of the SRF Cavities for the CEBAF 12 GeV Upgrade, *Proc. 15th Int. Conf. on RF Superconductivity,* Chicago, IL USA (2011), pp. 542-548,
http://accelconf.web.cern.ch/AccelConf/SRF2011/papers/tupo061.pdf.

[79]   F. Furuta, G. Hoffstaetter, M.Ge, M. Liepe, and B. Elmore, Multi-cell VEP Results: High Voltage,

High Q, and Localized Temperature Analysis, *Proc. Int. Part. Accel. Conf. 2012,* New Orleans, Louisiana (2012), pp. 1918-1920, http://accelconf.web.cern.ch/AccelConf/IPAC2012/papers/tuppr045.pdf.

[80] S. Calatroni, R. D. Waele, L. M. A. Ferreira, M. L. Macatrao, A. S. Skala, M. Sosin, and Y. L. Withofs, Status of the EP Simulations and Facilities for the SPL, *Proc. XXV Linear Accel. Conf.,* Tsukuba, Japan (2010), pp. 824-826, http://accelconf.web.cern.ch/AccelConf/LINAC2010/papers/thp032.pdf.

[81] F. Eozénou, S. Chel, Y. Gasser, J. P. Poupeau, C. Servouin, and Z. Wang, Vertical Electro-Polishing at CEA Saclay: Commissioning of a New Set-Up and Modeling of the Process Applied to Different Cavities, *Proc. 15th Int. Conf. on RF Superconductivity,* Chicago, IL USA (2011), pp. 549-554, http://accelconf.web.cern.ch/AccelConf/SRF2011/papers/tupo062.pdf.

[82] A. Dangwal Pandey, G. Müller, D. Reschke, and X. Singer, *Phys. Rev. ST Accel. Beams* **12**, 023501 (2009), http://link.aps.org/doi/10.1103/PhysRevSTAB.12.023501.

[83] C. E. Reece, E. Ciancio, K. A. Keyes, and D. Yang, A Study of the Effectiveness of Particulate Cleaning Protocols on Intentionally Contaminated Niobium Surfaces, *Proc. 14th Int. Conf. on RF Superconductivity,* Berlin, Germany (2009), pp. 746-750, http://accelconf.web.cern.ch/AccelConf/SRF2009/papers/thppo062.pdf.

[84] D. Sertore, et al., High pressure rinsing system studies, Beijing, China (2007), pp. 664-668, http://accelconf.web.cern.ch/accelconf/srf2007/PAPERS/TU102.pdf.

[85] D. Reschke, A. Brinkmann, K. Floettmann, D. Klinke, J. Ziegler, D. Werner, R. Grimme, and C. Zorn, Dry-ICE Cleaning: The Most Effective Cleaning Process for SRF Cavities?, *Proc. 13th Int. Workshop on RF Superconductivity,* Peking Univ., Beijing, China (2007), pp. 239-242, http://accelconf.web.cern.ch/AccelConf/srf2007/PAPERS/TUP48.pdf.

[86] D. Reschke and J. Ziegler, Open 120°C Bake in Argon Atmosphere: A Simplified Approach for Q-Drop Removal *Proc. XXIV Linear Accel. Conf.,* Victoria, BC, Canada (2008), pp. 809-811, http://accelconf.web.cern.ch/accelconf/LINAC08/papers/thp015.pdf.

[87] G. Ciovati, *Physica C: Superconductivity* **441**, 44 (2006), http://www.sciencedirect.com/science/article/pii/S0921453406001493.

[88] A. Gurevich and G. Ciovati, *Phys. Rev. B* **77**, 104501 (2008), http://link.aps.org/doi/10.1103/PhysRevB.77.104501.

[89] S. Aull, O. Kugeler, and J. Knobloch, *Phys. Rev. ST - Accel. and Beams* **15**, 062001 (2012), http://link.aps.org/doi/10.1103/PhysRevSTAB.15.062001.

[90] P. Dhakal, G. Ciovati, W. Rigby, J. Wallace, and G. R. Myneni, *Review of Scientific Instruments* **83**, 065105 (2012), http://dx.doi.org/10.1063/1.4725589.

[91] P. Brown, O. Brunner, A. Butterworth, E. Ciapala, H. Frischholz, G. Geschonke, E. Peschardt, and J. Sladen, Ultimate Performance of the LEP RF System, *Proc. 2001 Part. Accel. Conf.,* Chicago (2001), pp. 1059-1061, http://accelconf.web.cern.ch/AccelConf/p01/PAPERS/MPPH123.PDF.

[92] V. Arbet-Engels, C. Benvenuti, S. Calatroni, P. Darriulat, M. A. Peck, A. M. Valente, and C. A. Van't Hof, *NIM-A* **463**, 1 (2001), http://dx.doi.org/10.1016/S0168-9002(01)00165-6.

[93] C. Benvenuti, S. Calatroni, I. E. Campisi, P. Darriulat, M. A. Peck, R. Russo, and A. M. Valente, *Physica C* **316**, 153 (1999), http://dx.doi.org/10.1016/S0921-4534(99)00207-5.

[94] S. Stark, et al., A Novel Sputtered Medium Beta Cavity for ALPI, *Proc. 13th Int. Workshop on RF Superconductivity,* Peking Univ., Beijing, China (2007), pp. 216-218, http://accelconf.web.cern.ch/AccelConf/srf2007/PAPERS/TUP38.pdf.

[95] M. Therasse, O. Brunner, S. Calatroni, J. K. Chambrillon, B. Delaup, and M. Pasini, HIE-ISOLDE SRF Development Activities at CERN *Proc. 2nd Int. Part. Accel. Conf.,* San Sebastián, Spain (2011), pp. 316-318, http://accelconf.web.cern.ch/AccelConf/IPAC2011/papers/mopc104.pdf.

[96] M. Krishnan, et al., *Phys. Rev. ST Accel. Beams* **15**, 032001 (2012), http://link.aps.org/doi/10.1103/PhysRevSTAB.15.032001.

[97] X. Zhao, L. Phillips, C. E. Reece, Kang Seo, M. Krishnan, and E. Valderrama, *J. Appl. Phys.* **110**, 033523 (2011), http://dx.doi.org/10.1063/1.3611406

[98] J. K. Spradlin, O. Trofimova, and A.-M. Valente-Feliciano, Surface Preparation of Metallic Substrates for Quality SRF Thin Films, *Proc. 15th Int. Conf. on RF Superconductivity,* Chicago, IL USA (2011), pp. 936-939, http://accelconf.web.cern.ch/AccelConf/SRF2011/papers/thpo079.pdf.

[99] A.-M. Valente-Feliciano, Nb Films: Substrates, Nucleation & Crystal Growth, *Proc. 15th Int. Conf. on RF Superconductivity,* Chicago, IL USA (2011), pp. 332-342,

http://accelconf.web.cern.ch/AccelConf/SRF2011/
papers/tuiob06.pdf.

[100] J. K. Spradlin, H. L. Phillips, C. E. Reece, A.-M. Valente-Feliciano, X. Zhao, D. Gu, and K. I. Seo, Structural Properties of Niobium Thin Films Deposited on Metallic Substrates by ECR Plasma Energetic Condensation, *Proc. 15th Int. Conf. on RF Superconductivity,* Chicago, IL USA (2011), pp. 877-882, http://accelconf.web.cern.ch/AccelConf/SRF2011/papers/thpo064.pdf.

[101] X. Zhao, H. L. Phillips, C. E. Reece, J. K. Spradlin, A.-M. Valente-Feliciano, H. Baumgart, D. Gu, and K. I. Seo, Structural Characterization of Nb Films Deposited by ECR Plasma Energetic Condensation on Crystalline Insulators, *Proc. 15th Int. Conf. on RF Superconductivity,* Chicago, IL USA (2011), pp. 819-825, http://accelconf.web.cern.ch/AccelConf/SRF2011/papers/thpo044.pdf.

[102] C. Clavero, D. Beringer, R. A. Lukaszew, W. M. Roach, R. Skuza, and C. E. Reece, Strain Effects in the Superconducting Properties of Niobium Thin Films grown on Sapphire, *Proc. 15th Int. Conf. on RF Superconductivity,* Chicago, IL USA (2011), pp. 835-837, http://accelconf.web.cern.ch/AccelConf/SRF2011/papers/thpo047.pdf.

[103] W. M. Roach, D. Beringer, C. Clavero, R. A. Lukaszew, and C. E. Reece, Investigation of Epitaxial Niobium Thin Films Grown on Different Surfaces Suitable for SRF Cavities, *Proc. 15th Int. Conf. on RF Superconductivity,* Chicago, IL USA (2011), pp. 874-876, http://accelconf.web.cern.ch/AccelConf/SRF2011/papers/thpo062.pdf.

[104] D. Beringer, C. Clavero, R. A. Lukaszew, W. M. Roach, and C. E. Reece, Anomalous Morphological Scaling in Epitaxial Niobium Thin Films on MgO(001), *Proc. 15th Int. Conf. on RF Superconductivity,* Chicago, IL USA (2011), pp. 883-885, http://accelconf.web.cern.ch/AccelConf/SRF2011/papers/thpo065.pdf.

[105] W. M. Roach, D. B. Beringer, J. R. Skuza, W. A. Oliver, C. Clavero, C. E. Reece, and R. A. Lukaszew, *Phys. Rev. ST - Accel. and Beams* **15**, 062002 (2012), http://link.aps.org/doi/10.1103/PhysRevSTAB.15.062002.

[106] X. Zhao, et al., *Journal of Vacuum Science & Technology A* **27**, 620 (2009), http://dx.doi.org/10.1116/1.3131725.

[107] G. Müller, P. Kneisel, D. Mansen, H. Piel, J. Pouryamout, and R. Roeth, Nb3Sn Layers on High-Purity Nb Cavities with Very High Quality Factors and Accelerating Gradients, *Proc. 5th European Part. Accel. Conf.,* Sitges, Spain (1996),

pp. WEP002L, http://accelconf.web.cern.ch/AccelConf/e96/PAPERS/WEPL/WEP002L.PDF.

[108] S. Posen and M. Liepe, Stoichiometric Nb3Sn in First Samples Coated at Cornell, *Proc. 15th Int. Conf. on RF Superconductivity,* Chicago, IL USA (2011), pp. 886-889, http://accelconf.web.cern.ch/AccelConf/SRF2011/papers/thpo066.pdf.

[109] A.-M. Valente-Feliciano, H. L. Phillips, C. E. Reece, J. K. Spradlin, A. D. Batchelor, F. A. Stevie, R. A. Lukaszew, and K. I. Seo, SRF Multilayer Structures based on NbTiN, *Proc. 15th Int. Conf. on RF Superconductivity,* Chicago, IL USA (2011), pp. 920-925, http://accelconf.web.cern.ch/AccelConf/SRF2011/papers/thpo074.pdf.

[110] A.-M. Valente-Feliciano, H. L. Phillips, C. E. Reece, J. Spradlin, X. Zhao, and B. Xiao, Development of Nb and Alternative Material Thin Films Tailored for SRF Applications, *Proc. Int. Part. Accel. Conf. 2012,* New Orleans, Louisiana (2012), pp. 2444-2446, http://accelconf.web.cern.ch/AccelConf/IPAC2012/papers/weppc097.pdf.

[111] A. Gurevich and G. Ciovati, *Physical Review B* **77**, 104501 (2008),

[112] H. Padamsee, J. Knobloch, and T. Hays, *RF Superconductivity for Accelerators* (Wiley-VCH, 2008).

[113] G. Catelani and J. P. Sethna, *Phys. Rev. B* **78**, 224509 (2008),

[114] C. Buzea and T. Yamashita, *Supercond. Sci. Technol.* **14**, 115 (2001),

[115] E. Hand, *Nature* **456**, 555 (2008),

[116] A. Gurevich, *Applied Physics Letter* **88**, 012511 (2006),

[117] T. Tajima, et al., MgB2 Thin Film Studies, *Proc. 15th Int. Conf. on RF Superconductivity,* Chicago, IL USA (2011), pp. 287-292, http://accelconf.web.cern.ch/AccelConf/SRF2011/papers/tuioa04.pdf.

[118] X. X. Xi, *Supercond. Sci. Technol.* **22**, 043001 (2009), http://stacks.iop.org/0953-2048/22/i=4/a=043001.

[119] B. H. Moeckly and W. S. Ruby, *Supercond. Sci. Technol.* **19**, L21 (2006), http://stacks.iop.org/0953-2048/19/i=6/a=L02.

[120] B. P. Xiao, X. Zhao, J. Spradlin, C. E. Reece, M. J. Kelley, T. Tan, and X. X. Xi, *Supercond. Sci. Technol.* **25**, 095006 (2012), http://stacks.iop.org/0953-2048/25/i=9/a=095006.

[121] P. Bauer, N. Solyak, G. L. Ciovati, G. Eremeev, A. Gurevich, L. Lilje, and B. Visentin, *Physica C: Superconductivity* **441**, 51 (2006), http://www.sciencedirect.com/science/article/pii/S092145340600150X.

[122] G. Ciovati, P. Kneisel, and A. Gurevich, *Phys. Rev. ST Accel. Beams* **10**, 062002 (2007), http://link.aps.org/doi/10.1103/PhysRevSTAB.10.062002.

[123] S. Calatroni, *Physica C: Superconductivity* **441**, 95 (2006), http://www.sciencedirect.com/science/article/pii/S0921453406001596.

[124] T. Junginger, Ph.D. dissertation, Ruperto-Carola University of Heidelberg, 2012.

[125] J. Halbritter, Material science of Nb RF accelerator cavities: where do we stand 2001? , *Proc. 10th Int. Workshop on RF Superconductivity,* Tsukuba, Japan (2001), pp. 292-301, http://accelconf.web.cern.ch/accelconf/srf01/papers/ma006.pdf.

[126] G. Ciovati, *J. Appl. Phys.* **96**, 1591 (2004), http://dx.doi.org/10.1063/1.1767295.

[127] G. Ciovati and J. Halbritter, *Physica C: Superconductivity* **441**, 57 (2006), http://www.sciencedirect.com/science/article/pii/S0921453406001511.

[128] J. Amrit and C. Z. Antoine, *Phys. Rev. ST Accel. Beams* **13**, 023201 (2010), http://link.aps.org/doi/10.1103/PhysRevSTAB.13.023201.

[129] A. Gurevich, *Physica C: Superconductivity* **441**, 38 (2006), http://www.sciencedirect.com/science/article/pii/S0921453406001481.

[130] B. Visentin, Q-slope at high gradients: review of experiments and theories, Travemunde, Germany (2003), pp. 199-205, http://accelconf.web.cern.ch/accelconf/SRF2003/papers/tuo01.pdf.

[131] G. Ciovati, Review of high field Q-slope, cavity measurements, *Proc. 13th Int. Workshop on RF Superconductivity,* Beijing, China (2007), pp. 70-74, http://accelconf.web.cern.ch/accelconf/srf2007/PAPERS/TU102.pdf.

[132] G. Ciovati, G. Myneni, F. Stevie, P. Maheshwari, and D. Griffis, *Phys. Rev. ST Accel. Beams* **13**, 022002 (2010), http://link.aps.org/doi/10.1103/PhysRevSTAB.13.022002.

[133] W. Weingarten, *Phys. Rev. ST Accel. Beams* **14**, 101002 (2011), http://link.aps.org/doi/10.1103/PhysRevSTAB.14.101002.

[134] B. Visentin, M. F. Barthe, V. Moineau, and P. Desgardin, *Phys. Rev. ST Accel. Beams* **13**, 052002 (2010), http://link.aps.org/doi/10.1103/PhysRevSTAB.13.052002.

[135] A. Romanenko, A. Grassellino, J. P. Ozelis, and H. Padamsee, Depth distribution of losses in superconducting niobium cavities, *Proc. Int. Part. Accel. Conf. 2012,* New Orleans, LA (2012), pp. 2495-2497, http://accelconf.web.cern.ch/AccelConf/IPAC2012/papers/weppc116.pdf.

[136] B. Xiao, *Surface Impedance of Superconducting Radio Frequency Materials*, PhD dissertation, College of William and Mary, 2012.

[137] M. K. Transtrum, G. Catelani, and J. P. Sethna, *Phys. Rev. B* **83**, 094505 (2011), http://link.aps.org/doi/10.1103/PhysRevB.83.094505.

[138] G. Catelani and J. P. Sethna, *Phys. Rev. B* **78**, 224509 (2008), http://link.aps.org/doi/10.1103/PhysRevB.78.224509.

[139] F. P.-J. Lin and A. Gurevich, *Phys. Rev. B* **85**, 054513 (2012), http://link.aps.org/doi/10.1103/PhysRevB.85.054513.

[140] T. Yogi, G. J. Dick, and J. E. Mercereau, *Physical Review Letters* **39**, 826 (1977), http://link.aps.org/doi/10.1103/PhysRevLett.39.826.

[141] T. Hays and H. Padamsee, Measuring the RF critical field of Pb, Nb, and Nb3Sn, *Proc. 8th Int. Workshop on RF Superconductivity,* Abano Terme, Italy (1997), pp. 789-794, http://accelconf.web.cern.ch/accelconf/SRF97/papers/srf97d01.pdf.

[142] R. L. Geng, G. V. Eremeev, H. Padamsee, and V. D. Shemelin, High gradient studies for ILC with single cell re-entrant shape and elliptical shape cavities made of fine-grain and large-grain niobium, *Proc. 2007 Part. Accel. Conf.,* Albuquerque, NM (2007), pp. 2337-2339, http://accelconf.web.cern.ch/accelconf/p07/PAPERS/WEPMS006.PDF.

[143] C. Xu, H. Tian, C. E. Reece, and M. J. Kelley, *Phys. Rev. ST Accel. Beams* **14**, 123501 (2011), http://link.aps.org/doi/10.1103/PhysRevSTAB.14.123501.

[144] C. Xu, H. Tian, C. E. Reece, and M. J. Kelley, *Phys. Rev. ST Accel. Beams* **15**, 043502 (2012), http://link.aps.org/doi/10.1103/PhysRevSTAB.15.043502.

[145] R. L. Geng, J. Knobloch, and H. Padamsee, Microstructures of RF surfaces in the electron-beam-weld regions of niobium *Proc. 9th Int. Workshop on RF Superconductivity,* Santa Fe, NM (1999), pp. 238-245, http://accelconf.web.cern.ch/accelconf/SRF99/papers/tup021.pdf.

[146] M. Ge, G. Wu, D. Burk, J. Ozelis, E. Harms, D. Sergatskov, D. Hicks, and L. D. Cooley, *Supercond. Sci. Technol.* **24**, 035002 (8 pp) (2011), http://iopscience.iop.org/0953-2048/24/3/035002.

[147] J. Hao, D. Reschke, A. Brinkmann, and L. Lilje, Low temperature heat treatment effect on high-

field EP cavities *Proc. 11th Int. Workshop on RF Superconductivity,* Travemunde, Germany (2003), pp. 66-69, http://accelconf.web.cern.ch/accelconf/SRF2003/papers/mop16.pdf.

[148]   C. E. Reece, A. C. Crawford, and R. L. Geng, Improved Performance of JLab 7-Cell Cavities by Electropolishing, *Proc. 23rd Part. Accel. Conf.,* Vancouver, BC (2009), pp. 2126-2128, http://accelconf.web.cern.ch/AccelConf/PAC2009/papers/we5pfp055.pdf.

[149]   B. Visentin, J.P. Charrier, D. Roudier, Y. Gasser, A. Aspart, J. P. Poupeau, B. Coadou, and G. Monnereau, High gradient Q-slope: non in-situ baking, surface treatment by plasma and similarities between BCP & EP cavities, *Proc. 11th Int. Workshop on RF Superconductivity,* Travemunde, Germany (2003), pp. 74-77, http://accelconf.web.cern.ch/accelconf/SRF2003/papers/mop19.pdf.

[150]   P. Kneisel (private communication).

[151]   A. Dzyuba, A. Romanenko, and L. D. Cooley, *Supercond. Sci. Technol.* **23**, 125011 (2010), http://stacks.iop.org/0953-2048/23/i=12/a=125011.

[152]   C. Xu, M. J. Kelley, and C. E. Reece, Analysis of High Field Non-Linear Losses on SRF Surfaces Due to Specific Topographic Roughness, *Proc. Int. Part. Accel. Conf. 2012,* New Orleans, Louisiana (2012), pp. 2188-2190, http://accelconf.web.cern.ch/AccelConf/IPAC2012/papers/weeppb011.pdf.

[153]   P. Kneisel, S. Chattopadhyay, G. Ciovati, and G. Myneni, Performances of high-purity niobium cavities with different grain sizes, *Proc. 2006 Linear Accel. Conf.,* Knoxville, TN (2006), pp. 318-320, http://accelconf.web.cern.ch/accelconf/l06/PAPERS/TUP033.PDF.

[154]   J. Knobloch, R. L. Geng, M. Liepe, and H. Padamsee, High-field Q slope in superconducting cavities due to magnetic field enhancement at grain boundaries, *Proc. 9th Int. Workshop on RF Superconductivity,* Santa Fe, NM (1999), pp. 77-91, http://accelconf.web.cern.ch/accelconf/SRF99/papers/tua004.pdf.

[155]   A. A. Polyanskii, P. J. Lee, A. Gurevich, Z.-H. Sung, and D. C. Larbalestier, *AIP Conf. Proc.* **1352**, 186 (2011), http://dx.doi.org/10.1063/1.3579237.

[156]   Z. H. Sung, A. A. Polyanskii, P. J. Lee, A. Gurevich, and D. C. Larbalestier, *AIP Conf. Proc.* **1352**, 142 (2011), http://dx.doi.org/10.1063/1.3579232.

[157]   H. Safa, M. Boloré, Y. Boudigou, S. Jaidane, R. Keller, P. Nardin, and G. Szegedi, Specific resistance measurement of a single grain boundary in pure niobium, *Proc. 9th Int. Workshop on RF Superconductivity,* Santa Fe, NM (1999), pp. 267-269, http://accelconf.web.cern.ch/accelconf/SRF99/papers/tup029.pdf.

[158]   T. Proslier, J. F. Zasadzinski, L. Cooley, C. Antoine, J. Moore, J. Norem, M. Pellin, and K. E. Gray, *Appl. Phys. Lett.* **92**, 212505 (2008), http://dx.doi.org/10.1063/1.2913764.

[159]   J. Halbritter, *J. Appl. Phys.* **97**, 083904 (2005), http://dx.doi.org/10.1063/1.1874292.

[160]   R. Paul, H. H. Chen-Mayer, G. R. Myneni, W. A. Lanford, and R. E. Ricker, *Materiaux & Techniques* **91**, 23 (2003),

[161]   I. Arfaoui, C. Guillot, J. Cousty, and C. Antoine, *J. Appl. Phys.* **91**, 9319 (2002), http://dx.doi.org/10.1063/1.1473699.

[162]   M. Grundner and J. Halbritter, *J. Appl. Phys.* **51**, 397 (1980), http://dx.doi.org/10.1063/1.327386.

[163]   B. Visentin, J.-P. Charrier, Y. Gasser, and S. Regnaud, Fast Argon-baking process for mass production of niobium superconducting RF cavities, *Proc. 2006 European Part. Accel. Conf.,* Edinburgh, Scotland (2006), pp. 381-383, http://accelconf.web.cern.ch/accelconf/e06/PAPERS/MOPCH141.PDF.

[164]   M. Delheousy, *X-ray investigation of Nb/O interfaces,* University of Paris-Sud IX and Stuttgart University, 2008.

[165]   J. Halbritter, *Applied Physics A: Materials Science & Processing* **43**, 1 (1987), http://dx.doi.org/10.1007/BF00615201.

[166]   H. Tian, B. Xiao, M. J. Kelley, C. E. Reece, A. DeMasi, L. Pipe, and K. E. Smith, Recent XPS studies of the effect of processing on Nb SRF surfaces, *Proc. 13th Int. Workshop on RF Superconductivity,* Beijing, China (2007), pp. 158-162, http://accelconf.web.cern.ch/accelconf/srf2007/PAPERS/TUP18.pdf.

[167]   K. E. Yoon, D. N. Seidman, C. Antoine, and P. Bauer, *Appl. Phys. Lett.* **93**, 132502 (2008), http://dx.doi.org/10.1063/1.2987483.

[168]   Q. Ma, P. Ryan, J. W. Freeland, and R. A. Rosenberg, *J. Appl. Phys.* **96**, 7675 (2004), http://dx.doi.org/10.1063/1.1809774.

[169]   M. Delheusy, et al., *Appl. Phys. Lett.* **92**, 101911 (2008), http://dx.doi.org/10.1063/1.2889474.

[170]   H. Tian, M. J. Kelley, C. E. Reece, S. Wang, L. Plucinski, K. E. Smith, and M. Nowell, *Appl. Surf. Sci.* **253**, 1236 (2006),

[171]   F. Palmer, Surface resistance of superconductors - examples from Nb-O systems Argonne, IL (1987), pp. 309-330, http://accelconf.web.cern.ch/accelconf/srf87/papers/srf87c06.pdf.

[172]   T. Proslier, M. Kharitonov, M. Pellin, J. Zasadzinski, and G. Ciovati, *IEEE Trans. Appl.*

*Superconductivity* **21**, 2619 (2011),
http://dx.doi.org/10.1109/TASC.2011.2107491.

[173] C. Z. Antoine, S. Berry, S. Bouat, J. F. Jacquot, J.
C. Villegier, G. Lamura, and A. Gurevich, *Phys.
Rev. ST Accel. Beams* **13**, 121001 (2010),
http://link.aps.org/doi/10.1103/PhysRevSTAB.13.1
21001.

[174] C. Z. Antoine, S. Berry, M. Aurino, J. Jacquot, J.
Villegier, G. Lamura, and A. Andreone, *IEEE
Trans. Appl. Superconductivity* **21**, 2601 (2011),
http://dx.doi.org/10.1109/TASC.2010.2100347.

[175] T. Proslier, et al., Atomic Layer Deposition for
SRF Cavities, *Proc. 23rd Part. Accel. Conf.,*
Vancouver, BC (2009), pp. 803-805,
http://accelconf.web.cern.ch/AccelConf/PAC2009/
papers/tu5pfp002.pdf.

[176] T. Proslier, J. Klug, N. C. Becker, J. W. Elam, and
M. Pellin, *ECS Transactions* **41**, 237 (2011),
http://dx.doi.org/10.1149/1.3633673.

[177] A 2-cell temperature mapping system for ILC
cavities, 2008,

[178] D. F. Orris, et al., Fast Thermometry for
Superconducting RF Cavity Testing, *Proc. 2007
Part. Accel. Conf.,* Albuquerque, NM (2007), pp.
2280-2282,
http://accelconf.web.cern.ch/accelconf/p07/PAPER
S/WEPMN105.PDF.

[179] A. Canabal, F. L. Krawczyk, R. J. Roybal, J. D.
Sedillo, T. Tajima, S. Cohen, and W. Haynes, Full
real-time temperature mapping system for 9-cell
ILC-type cavities, *Proc. 2008 European Part.
Accel. Conf.,* Genoa, Italy (2008), pp. 841-843,
http://accelconf.web.cern.ch/accelconf/e08/papers/
mopp121.pdf.

[180] K. Liao, L. Arnaudon, O. Brunner, E. Ciapala, D.
Glenat, and W. Weingarten, Design and
Development of an Octopus Thermometric System
for the 704 MHz Single-cell SPL Cavity at CERN,
*Proc. Int. Part. Accel. Conf. 2012,* New Orleans,
Louisiana (2012), pp. 2266-2268,
http://accelconf.web.cern.ch/AccelConf/IPAC2012
/papers/weppc029.pdf.

[181] Z. A. Conway, D. L. Hartill, H. Padamsee, and E.
N. Smith, Defect location in superconducting
cavities cooled with He-II using oscillating
superleak transducers *Proc. 14th Int. Conf. on RF
Superconductivity,* Berlin, Germany (2009), pp.
113-116,
http://accelconf.web.cern.ch/accelconf/SRF2009/p
apers/tuoaau05.pdf.

[182] Y. Iwashita, Y. Tajima, and H. Hayano, *Phys. Rev.
ST Accel. Beams* **11**, 093501 (2008),
http://link.aps.org/doi/10.1103/PhysRevSTAB.11.0
93501.

[183] R. L. Geng and T. Goodman, A Machine for High-
Resolution Inspection of SRF Cavities at JLab,
*Proc. 15th Int. Conf. on RF Superconductivity,*
Chicago, IL USA (2011), pp. 798-800,

http://accelconf.web.cern.ch/AccelConf/SRF2011/
papers/thpo036.pdf.

[184] K. Watanabe, H. Hayano, E. Kako, S. Noguchi, T.
Shishido, Y. Yamamoto, and Y. Iwashita, Repair
techniques of superconducting cavity for
improvement cavity performance at KEK-STF,
*Proc. 1st Int. Part. Accel. Conf.,* Kyoto, Japan
(2010), pp. 2965-2967,
http://accelconf.web.cern.ch/accelconf/IPAC10/pa
pers/wepec033.pdf.

[185] G. Ciovati, et al., *Rev. Sci. Instrum.* **83**, 034704
(2012), http://dx.doi.org/10.1063/1.3694570.

[186] T. Junginger, W. Weingarten, and C. Welsch,
*Review of Scientific Instruments* **83**, 063902
(2012), http://dx.doi.org/10.1063/1.4725521.

[187] B. P. Xiao, C. E. Reece, H. L. Phillips, R. L. Geng,
H. Wang, F. Marhauser, and M. J. Kelley, *Rev. Sci
.Instrum.* **82**, 056104 (2011),
http://link.aip.org/link/doi/10.1063/1.3575589.

[188] Y. Xie and M. Liepe, TE Sample Host Cavities
Development at Cornell, *Proc. 15th Int. Conf. on
RF Superconductivity,* Chicago, IL USA (2011),
pp. 841-845,
http://accelconf.web.cern.ch/AccelConf/SRF2011/
papers/thpo050.pdf.

[189] J. Guo, D. W. Martin, S. G. Tantawi, and C.
Yoneda, Testing the RF Properties of Novel Super
Conducting Materials, *Proc. 15th Int. Conf. on RF
Superconductivity,* Chicago, IL (2011), pp. 325-
329,
http://accelconf.web.cern.ch/AccelConf/SRF2011/
papers/tuiob03.pdf.

[190] N. Pogue, P. M. McIntyre, and C. E. Reece,
Ultimate-Gradient Srf Test Cavity and Low Loss
Tangent Measurements in Ultra Pure Sapphire,
*Proc. XXV Linear Accel. Conf.,* Tsukuba, Japan
(2010), pp. 842-844,
http://accelconf.web.cern.ch/AccelConf/LINAC20
10/papers/thp038.pdf.

[191] S. Posen and M. Liepe, *Phys. Rev. ST Accel.
Beams* **15**, 022002 (2012),
http://link.aps.org/doi/10.1103/PhysRevSTAB.15.0
22002.

[192] L. Xiao, K. Ko, K. H. Lee, C.-K. Ng, M. Liepe,
and N. R. A. Valles, Effects of Elliptically
Deformed Cell Shape in the Cornell ERL Cavity,
*Proc. 15th Int. Conf. on RF Superconductivity,*
Chicago, IL USA (2011), pp. 244-246,
http://accelconf.web.cern.ch/AccelConf/SRF2011/
papers/mopo061.pdf.

[193] F. Marhauser, J. Henry, and H. Wang, Critical
Dipole Modes in JLAB Upgrade Cavities, *Proc.
XXV Linear Accel. Conf.,* Tsukuba, Japan (2010),
pp. 776-778,
http://accelconf.web.cern.ch/AccelConf/LINAC20
10/papers/thp009.pdf.

[194] *International Workshop on Higher-Order-Mode
Diagnostics and Suppression in Superconducting*

*Cavities* (Cockcroft Institute, Daresbury, UK, 2012) http://www.cockcroft.ac.uk/events/HOMSC12/.

[195] *Accelerators for America's Future* (US Dept. of Energy, 2012) http://www.acceleratorsamerica.org/.

[196] M. Lindroos, R. Ainsworth, R. D. Duperrier, S. Molloy, S. Peggs, K. Rathsman, R. Zeng, and R. J. M. Y. Rube, The ESS Accelerator, *Proc. 15th Int. Conf. on RF Superconductivity,* Chicago, IL USA (2011), pp. 994-998, http://accelconf.web.cern.ch/AccelConf/SRF2011/papers/friob06.pdf.

[197] N. Nakamura, Review of ERL Projects at KEK and Around the World, *Proc. Int. Part. Accel. Conf. 2012,* New Orleans, Louisiana, USA (2012), pp. 1040-1044, http://accelconf.web.cern.ch/AccelConf/IPAC2012/papers/tuxb02.pdf.

[198] M. Liepe, Y. He, G. H. Hoffstaetter, S. Posen, J. Sears, V. D. Shemelin, M. Tigner, N. R. A. Valles, and V. Veshcherevich, Superconducting RF for the Cornell Energy-Recovery Linac Main Linac, *Proc. 15th Int. Conf. on RF Superconductivity,* Chicago, IL USA (2011), pp. 90-96, http://accelconf.web.cern.ch/AccelConf/SRF2011/papers/mopo016.pdf.

[199] S. Sakanaka, H. Kawata, Y. Kobayashi, N. Nakamura, and R. Hajima, Construction Status of the Compact ERL, *Proc. Int. Part. Accel. Conf. 2012,* New Orleans, Louisiana (2012), pp. 607-609, http://accelconf.web.cern.ch/AccelConf/IPAC2012/papers/moppp018.pdf.

[200] P. A. McIntosh, C. D. Beard, and D. M. Dykes, SRF Linac Solutions for 4GLS at Daresbury, *Proc. 2006 Linear Accel. Conf.,* Knoxville, TN (2006), pp. 64-66, http://accelconf.web.cern.ch/AccelConf/l06/PAPERS/MOP016.PDF.

[201] Y. Yamamoto, et al., Beam Commissioning Status of Superconducting Crab Cavities in KEKB, *Proc. 1st Int. Part. Accel. Conf.,* Kyoto, Japan (2010), pp. 42-44, http://accelconf.web.cern.ch/AccelConf/IPAC10/papers/moocmh03.pdf.

[202] H. Wang, et al., Crab Cavity and Cryomodule Prototype Development for the Advanced Photon Source, *Proc. 2011 Part. Accel. Conf.,* New York, USA (2011), pp. 1472-1474, http://accelconf.web.cern.ch/AccelConf/PAC2011/papers/weocs7.pdf.

[203] *LHC-CC11, 5th LHC Crab Cavity Workshop* (CERN, Geneva, 2011) https://indico.cern.ch/conferenceDisplay.py?confId=149614.

[204] J. D. Fuerst, D. Horan, J. Kaluzny, A. Nassiri, T. L. Smith, and G. Wu, Tests of SRF Deflecting Cavities at 2K, *Proc. Int. Part. Accel. Conf. 2012,* New Orleans, Louisiana (2012), pp. 2300-2302, http://accelconf.web.cern.ch/AccelConf/IPAC2012/papers/weppc041.pdf.

[205] J. R. Delayen, Compact Superconducting Cavities for Deflecting and Crabbing Applications *Proc. 15th Int. Conf. on RF Superconductivity,* Chicago, IL USA (2011), pp. 631-636, http://accelconf.web.cern.ch/AccelConf/SRF2011/papers/thioa03.pdf.

[206] A. Arnold and J. Teichert, *Phys. Rev. ST Accel. Beams* **14**, 024801 (2011), http://link.aps.org/doi/10.1103/PhysRevSTAB.14.024801.

[207] A. Neumann, et al., SRF Photoinjector Tests at HoBiCat, *Proc. 15th Int. Conf. on RF Superconductivity,* Chicago, IL USA (2011), http://accelconf.web.cern.ch/AccelConf/SRF2011/papers/frioa07.pdf.

[208] P. Murcek, et al., Modified SRF Photoinjector for the ELBE at HZDR, *Proc. 15th Int. Conf. on RF Superconductivity,* Chicago, IL USA (2011), http://accelconf.web.cern.ch/AccelConf/SRF2011/papers/mopo004.pdf.

[209] F. Zhu, et al., Status of the DC-SRF Photoinjector for PKU-SETF, *Proc. 15th Int. Conf. on RF Superconductivity,* Chicago, IL USA (2011), http://accelconf.web.cern.ch/AccelConf/SRF2011/papers/friob02.pdf.

[210] R. Legg, et al., Status of the Wisconsin SRF Gun, *Proc. Int. Part. Accel. Conf. 2012,* New Orleans, Louisiana, USA (2012), pp. 661-663, http://accelconf.web.cern.ch/AccelConf/IPAC2012/papers/moppp045.pdf.

[211] J. N. Corlett, et al., Next Generation Light Source R&D and Design Studies at LBNL, *Proc. Int. Part. Accel. Conf. 2012,* New Orleans, Louisiana (2012), pp. 1762-1764, http://accelconf.web.cern.ch/AccelConf/IPAC2012/papers/tuppp070.pdf.

[212] R. W. Garnett, L. J. Rybarcyk, D. E. Rees, and T. Tajima, High Power Options for LANSCE, *Proc. 2011 Part. Accel. Conf.,* New York, USA (2011), pp. 2107-2110, http://accelconf.web.cern.ch/AccelConf/PAC2011/papers/thocn4.pdf.

[213] R. W. Garnett, F. E. Merrill, J. G. O'Hara, D. E. Rees, L. J. Rybarcyk, T. Tajima, and P. L. Walstrom, A Conceptual 3-GeV Linac Upgrade for Enhanced Proton Radiography, *Proc. Int. Part. Accel. Conf. 2012,* New Orleans, Louisiana, USA (2012), pp. 4130-4133, http://accelconf.web.cern.ch/AccelConf/IPAC2012/papers/thppr067.pdf.

[214] W. S. Graves, W. Brown, F. X. Kaertner, and D. E. Moncton, *Nuclear Instruments and Methods in Physics Research Section A: Accelerators, Spectrometers, Detectors and Associated*

*Equipment* **608**, S103 (2009),
http://www.sciencedirect.com/science/article/pii/S
0168900209009802.

[215] G. A. Krafft and G. Priebe, *Reviews of Accelerator Science and Technology* **3**, 147 (2010),
http://www.worldscinet.com/rast/03/0301/S179362
6810000440.html.

[216] E. I. Simakov, W. B. Haynes, M. A. Madrid, F. P. Romero, T. Tajima, W. M. Tuzel, C. H. Boulware, and T. L. Grimm, An Update on a Superconducting Photonic Band Gap Structure Resonator Experiment, *Proc. Int. Part. Accel. Conf. 2012,* New Orleans, Louisiana, USA (2012), pp. 2140-2142,
http://accelconf.web.cern.ch/AccelConf/IPAC2012
/papers/weoab03.pdf.

[217] E. I. Simakov, C. H. Boulware, and T. L. Grimm, Design of a Superconducting Photonic Band Gap Structure Cell, *Proc. 2011 Part. Accel. Conf.,* New York, USA (2011), pp. 178-181,
http://accelconf.web.cern.ch/AccelConf/PAC2011/
papers/mop042.pdf.

[218] D. Vandeplassche and L. M. Romão, Accelerator Driven Systems, *Proc. Int. Part. Accel. Conf. 2012,* New Orleans, Louisiana (2012), pp. 6-10,
http://accelconf.web.cern.ch/AccelConf/IPAC2012
/papers/moyap01.pdf.

[219] P. Singh, SRF Accelerator for Indian ADS Program: Present & Future Prospects, *Proc. 15th Int. Conf. on RF Superconductivity,* Chicago, IL USA (2011), pp. 983-987,
http://accelconf.web.cern.ch/AccelConf/SRF2011/
papers/friob04.pdf.

[220] S. Fu, et al., Chinese Plan for ADS and CSNS, *Proc. 15th Int. Conf. on RF Superconductivity,* Chicago, IL (2011), pp. 977-982,
http://accelconf.web.cern.ch/AccelConf/SRF2011/
papers/friob03.pdf.

[221] D. Vandeplassche, J.-L. Biarrotte, H. Klein, and H. Podlech, The Myrrha Linear Accelerator, *Proc. 2nd Int. Part. Accel. Conf.,* San Sebastián, Spain (2011), pp. 2718-2720,
http://accelconf.web.cern.ch/AccelConf/IPAC2011
/papers/weps090.pdf.

[222] *Accelerator and Target Technology for Accelerator Driven Transmutation and Energy Production* (FNAL, 2010) FERMILAB-FN-0907-DI, http://inspirehep.net/record/873011/.

[223] Project X –Reference Design Report, Version 1.0, 2010, S. D. Holmes ed., FNAL, http://projectx-docdb.fnal.gov/cgi-bin/RetrieveFile?docid=776;filename=RDR-v0.9.pdf;version=1.

[224] O. S. Brüning, F. Zimmermann, and M. Klein, The LHeC Project Development Beyond 2012, *Proc. Int. Part. Accel. Conf. 2012,* New Orleans, Louisiana (2012), pp. 1999-2001,
http://accelconf.web.cern.ch/AccelConf/IPAC2012
/papers/tuppr076.pdf.

[225] F. Gerigk, et al., Layout and Machine Optimisation for the SPL at CERN, *Proc. XXV Linear Accel. Conf.,* Tsukuba, Japan (2010), pp. 761-763,
http://accelconf.web.cern.ch/AccelConf/LINAC20
10/papers/thp004.pdf.

[226] A. P. Blondel, F. Zimmermann, M. Koratzinos, and M. Zanetti, LEP3: A High Luminosity e+e-Collider in the LHC Tunnel to Study the Higgs Boson, *Proc. Int. Part. Accel. Conf. 2012,* New Orleans, Louisiana (2012), pp. 2005-2007,
http://accelconf.web.cern.ch/AccelConf/IPAC2012
/papers/tuppr078.pdf.

[227] H. Hayano, Progress and Plans for RD and the Conceptual Design of the ILC Main Linacs, *Proc. 2005 Part. Accel. Conf.,* Knoxville (2005), pp. 199-203,
http://accelconf.web.cern.ch/AccelConf/p05/PAPE
RS/WOAA002.PDF.

[228] A. Yamamoto, M. C. Ross, and N. J. Walker, Advances in SRF Development for ILC, *Proc. 15th Int. Conf. on RF Superconductivity,* Chicago, IL USA (2011), pp. 7-12,
http://accelconf.web.cern.ch/AccelConf/SRF2011/
papers/moioa02.pdf.

[229] S. Geer, Muon Colliders and Neutrino Factories, *Proc. XXV Linear Accel. Conf.,* Tsukuba, Japan (2010), pp. 1048-1052,
http://accelconf.web.cern.ch/AccelConf/LINAC20
10/papers/fr202.pdf.

[230] M. S. Zisman, R&D Toward a Neutrino Factory and Muon Collider, *Proc. 2011 Part. Accel. Conf.,* New York, USA (2011), pp. 2056-2060,
http://accelconf.web.cern.ch/AccelConf/PAC2011/
papers/thobn1.pdf.

[231] R. L. Geng, P. Barnes, D. Hartill, H. Padamsee, J. Sears, S. Calatroni, E. Chiaveri, R. Losito, and H. Preis, 200 MHz Nb-Cu Cavities for Muon Acceleration, *Proc. 11th Int. Workshop on RF Superconductivity,* Travemunde, Germany (2003), pp. 531-534,
http://accelconf.web.cern.ch/AccelConf/SRF2003/
papers/weo06.pdf.

[232] http://accelconf.web.cern.ch/accelconf/.